\documentclass{aastex}
\usepackage{emulateapj5}
\usepackage{epsfig}

\newcommand{\thalf}{t_{1\!/2}}
\newcommand{\tE}{t_{\rm E}}

\newcommand{\qmax}{Q_{\rm max}}
\newcommand{\Dfm}{\Delta F_{\rm max}}

\begin{document}
\title{Microlensing Candidates in M87 and the Virgo Cluster\\ with the Hubble
Space Telescope}
\author{Edward~A.~Baltz$^1$, Tod~R.~Lauer$^2$, David~R.~Zurek$^3$
Paolo~Gondolo$^4$, Michael~M.~Shara$^3$, Joseph~Silk$^5$, Stephen~E.~Zepf$^6$}

\affil{1.\ KIPAC, Stanford University,
P.O. Box 90450, MS 29, Stanford, CA 94309, {\tt eabaltz@slac.stanford.edu}\\
2.\ National Optical Astronomy Observatory, P.O. Box 26732, Tucson, AZ 85726 \\
3.\ American Museum of Natural History, Central Park West at 79th Street, New
York, NY 10024\\
4.\ Physics Department, University of Utah, 115 S 1400 E, Suite 201, Salt Lake
City, UT 84112\\
5.\ University of Oxford, Astrophysics, Keble Road, Oxford, OX1 3RH, United
Kingdom\\
6.\ Department of Physics and Astronomy, Michigan State University, East
Lansing, MI 48824}

\begin{abstract}
\end{abstract}

The position of the giant elliptical galaxy M87 at the center of the Virgo
Cluster means that the inferred column density of dark matter associated with
both the cluster halo and the galaxy halo is quite large.  This system is thus
an important laboratory for studying massive dark objects in elliptical
galaxies and galaxy clusters by gravitational microlensing, strongly
complementing the studies of spiral galaxy halos performed in the Local Group.
We have performed a microlensing survey of M87 with the WFPC2 instrument on the
Hubble Space Telescope.  Over a period of thirty days, with images taken once
daily, we discover seven variable sources.  Four are variable stars of some
sort, two are consistent with classical novae, and one exhibits an excellent
microlensing lightcurve, though with a very blue color implying the somewhat
disfavored possibility of a horizontal branch source being lensed.  Based on
sensitivity calculations from artificial stars and from artificial lightcurves,
we estimate the expected microlensing rate.  We find that the detection of one
event is consistent with a dark halo with a 20\% contribution of microlensing
objects for both M87 and the Virgo Cluster, similar to the value found from
observations in the Local Group.  Further work is required to test the
hypothesized microlensing component to the cluster.

\keywords{gravitational lensing --- galaxies: clusters: individual (Virgo) ---
  galaxies: halos --- galaxies: individual (M87) --- dark matter}

\section{Introduction}

In a classic paper Paczy\'nski (1986) proposed a search for massive dark
objects in the Milky Way halo by searching for the rare gravitational
microlensing of Large and Small Magellanic Cloud (LMC and SMC) stars.  For a
halo consisting of roughly solar--mass objects, of order one in one million LMC
stars is being lensed (with a magnification of 30\% or more) at any given time.
An extensive monitoring campaign could thus hope to detect these transient
lensing events, which develop over typically 100 days, thus elucidating the
nature of the Milky Way halo.  This has been accomplished with great success by
several groups, described below.  Furthermore, the extension of this work to
other nearby galaxies is well underway.

The MACHO project (Alcock et al.~2000) monitored the LMC for microlensing
events for the better part of a decade: they conclude that there is an excess
of events over the expectation from known stellar populations corresponding to
an approximately 20\% contribution of sub-solar-mass objects to the dark halo
of the Milky Way.  The EROS collaboration (Afonso et al.~2003) has monitored
the LMC and SMC over a similar time period, and finds only an upper limit of
25\% on the microlensing component.

As proposed a decade ago (Crotts 1992), the Andromeda Galaxy (M31) is an
excellent target for a microlensing survey.  Both the Milky Way and M31 halos
can be studied in detail.  Very few stars are resolved from the ground, thus
image subtraction is required.  This is the ``pixel'' lensing regime (Crotts
1992; Baillon et al.~1993; Gould 1996).  Several collaborations, including
MEGA (preceded by the VATT/Columbia survey), AGAPE, and WeCAPP, have produced a
number of microlensing event candidates involving stars in M31 (Crotts \&
Tomaney 1996, Ansari et al.~1999, Auri\`ere et al.~2001, Uglesich 2001, Calchi
Novati et al.~2002, de Jong et al.~2003, Riffeser et al.~2003).  The results of
the VATT/Columbia survey (Uglesich et al.~2003) are inconclusive, possibly
indicating the presence of a microlensing halo of sub-solar-mass objects around
M31.

Finally, we turn to the subject of this paper, the giant elliptical galaxy M87
in the Virgo cluster.  The ability of the Hubble Space Telescope (HST) to
perform a microlensing survey of the Virgo cluster was noted several years ago
(Gould 1995).  A variability survey of M87 could discover a microlensing
population either in the M87 halo or even an intracluster population in the
overall Virgo halo.  We have used thirty orbits of HST data from the WFPC2 to
perform just such a survey.

All of these observational programs are aimed at understanding the nature of
the dark halos of galaxies.  The halos of the large spiral galaxies of the
local group (the Milky Way and M31) can be studied from the ground.  M87 is a
particularly interesting target because it is an elliptical galaxy, and as such
contains a different population of stars.  Furthermore, it serves to
``illuminate'' any dark objects in the halo of the Virgo cluster.  Such objects
might have been stripped from their host galaxies in the formation of Virgo, as
the tidal effects during galaxy mergers and other interactions should have been
substantial.

The outline of this paper is as follows.  We briefly discuss the theory of
microlensing in \S\ref{sec:theory}.  Our HST observations of M87, data
reduction, and image subtraction and filtering are covered in
\S\ref{sec:imageanalysis}.  Selection of candidate events, including the
exclusion of hot pixels is discussed in \S\ref{sec:candidateselection}.  We
describe the variable source detection efficiency in
\S\ref{sec:detectionefficiency}, and the calculation of the microlensing rate
in \S\ref{sec:modeling}.  We conclude with a discussion in
\S\ref{sec:discussion}.

\section{Theory of Microlensing}
\label{sec:theory}
The term microlensing refers to the fact that the multiple images of source
stars are split by microarcseconds.  The splitting is unobserved, but the
magnification can be large.  It is the transient magnification that is sought.

\subsection{Basics}

We now lay out the basic physics and terminology of gravitational microlensing.
For a point mass, the lens equation can be written in terms of the Einstein
radius and angle, given by
\begin{equation}
R_{\rm E}=\sqrt{\frac{4GM}{c^2}\frac{D_{\rm l}D_{\rm ls}}{D_{\rm s}}},\;\;\;
\theta_{\rm E}=\sqrt{\frac{4GM}{c^2}\frac{D_{\rm ls}}{D_{\rm s}D_{\rm l}}},
\end{equation}
where $M$ is the lens mass, $D_{\rm l}$ is the distance to the lens, $D_{\rm
s}$ is the distance to the source, and $D_{\rm ls}$ is the distance between the
lens and source.  In angular coordinates, the lens equation relating source
position $\theta_S$ and image position $\theta_I$ is then
\begin{equation}
\vec{\theta}_S=\vec{\theta}_I\left(1-\frac{\theta_{\rm
E}^2}{\theta_I^2}\right).
\end{equation}
There are always two images.  This equation is usually written with all angles
written in units of the Einstein angle.  In particular $u=\theta_S/\theta_{\rm
E}$.  The magnification was first given by Einstein (1936):
\begin{equation}
A(u)=1+f(u^2)=1+\delta(u),\;\;\;f(x)=\frac{2+x}{\sqrt{x(4+x)}}-1.
\end{equation}
For $u\ll1$, $A\approx 1/u$, namely the magnification can be very large.

The timescale over which a microlensing event progresses is the Einstein time
$\tE=R_{\rm E}/v_\perp$, given in terms of the Einstein radius and the
perpendicular velocity of the lens relative to the line of sight to the source
$v_\perp$.  Assuming rectilinear motion for the lens, we find
\begin{equation}
u^2(t)=\beta^2+\frac{(t-t_0)^2}{\tE^2},
\end{equation}
where $\beta$ is the minimum impact parameter of the lens.  For a star with
unlensed flux $F_\star$, the microlensing lightcurve is then
\begin{equation}
F(t)=F_\star+F_\star\delta(u(t)).
\end{equation}
If $F_\star$ can be measured, the Einstein time can be extracted from the event
lightcurve.  The shape of the lightcurve does contain some information on
$\beta$ (and thus $F_\star$), though extracting it requires very high quality
data (e.g.\ Paulin-Henriksson et al.~2003).

This discussion has assumed that the angular size of the source
$(\theta_\star=R_\star/D_{\rm l})$ is much smaller than the minimum impact
parameter in Einstein units, namely $\theta_\star\ll\beta\theta_{\rm E}$.  If
this is not the case, finite source size effects can be significant, as
summarized by Yoo et al.~(2004).

\subsection{``Pixel'' Microlensing}
Unfortunately, the unlensed flux of the source star is not easy to measure.
Even nearby sources (e.g.\ in the LMC) may be significantly blended.  For
sources in M87, the blending is always severe, and the unlensed flux is all but
unmeasurable.  The Einstein timescale is notoriously hard to determine for
such events.  The measured timescale is in effect the full width at half
maximum timescale, given by $\thalf=2\,\tE\,w(\beta)$, where
\begin{equation}
w(\beta)=\sqrt{2f(f(\beta^2))-\beta^2},
\end{equation}
with limiting behavior
\begin{equation}
w(\beta\ll1)=\beta\sqrt{3},\;\;\;w(\beta\gg1)=\beta\sqrt{\sqrt{2}-1}.
\end{equation}
For all values of $\beta$ we thus find that $\thalf\sim \tE\,\beta$.  For
the present work, high magnifications are required, implying small values of
$\beta$, and thus full width at half maximum timescales much smaller than the
Einstein timescales.

In the high magnification limit $\beta\rightarrow0$, we can write down the
``degenerate'' form of the microlensing lightcurve,
\begin{eqnarray}
F(t)&=&B+\Dfm\left[1+12\left(\frac{t-t_0}{\thalf}\right)^2\right]^{-1/2},
\label{eq:lc1}\\ &=&B+\Dfm\,\left(\frac{\beta}{u(t)}\right), \label{eq:lc2}
\end{eqnarray}
where $B$ is the baseline flux and $\Dfm=F_\star\delta(\beta)\approx
F_\star/\beta$ is a fit parameter expressing the maximum {\em increase} in flux
from the lensed star.  In the absence of blending $B=F_\star$, but for M87 in
practice blending is significant.  This is the ``pixel'' lensing regime in
which sources are only resolved when they are lensed.  The measured parameters
are $\Dfm$ and $\thalf$.  As mentioned previously, the Einstein time $\tE$ can
not be easily measured without measuring $F_\star$, though for high
signal-to-noise data (Paulin-Henriksson et al.~2003) this problem is
ameliorated.  We can proceed however, though a distribution in $\thalf$ for
microlensing events is not as useful as a distribution in $\tE$.  In the high
magnification regime, we note that $\Dfm\thalf\sim F_\star \tE$ (though this
relation is affected by finite source size effects), so we can recover Einstein
times statistically at least, given that the distribution of $F_\star$ is known
(Gondolo 1999).  This distribution is simply the stellar luminosity function.
Stellar population models can give a reasonable estimate for the luminosity
function, and the first non-trivial moment is known: this is just the surface
brightness fluctuation (SBF) magnitude.

\section{Observations and Image Analysis}
\label{sec:imageanalysis}

\subsection{Design of the Observational Program}

Our {\it HST} microlensing program, GO-8592, was awarded 30 orbits for WFPC2
imagery in May/June 2001; a journal of observations is presented in
Table~\ref{tab:journal}.  This allocation comprised a program of daily
single-orbit sampling of M87 over a month-long interval.  Within each orbit we
obtained four 260s exposures in the $I$-band F814W filter, yielding a 1040s
total exposure, followed by a single 400s broad-R F606W exposure to obtain
color information for any variables identified.  The four F814W pictures were
dithered by steps of 0.5 WFC-pixels aligned with the CCD axes, which in the
$I$-band allows a Nyquist-sampled interlace image to be constructed for the WFC
CCDs (Lauer 1999a); the PC1 CCD is already critically-sampled at F814W.  In
passing, the sharper PSF at F606W would require a $3\times3$ dither pattern to
lift the aliasing.  This schema establishes the F814W frames as the primary
search imagery, relegating the F606W to providing auxiliary color information
and verification of the events.  To obtain equal quality in F606W (or another
bluer filter) would be prohibitive; microlensed stars are expected to be red,
requiring more than a double allocation of orbits.  The alternative of
splitting a single orbit equally between two filters would reduce the depth in
any single filter, and would make dithering prohibitively expensive.  While the
dithering scheme that we did adopt does include an overhead that might
otherwise be used to collect photons, this exposure-time penalty is more than
offset by the resolution gain returned by Nyquist sampling.  A simpler strategy
of obtaining single, or CR-split identical exposures, in each filter within an
orbit would reduce the sensitivity to detecting faint point-sources against the
M87 envelope.  The nucleus of M87 was centered in PC1.  The pointing and
orientation were held fixed over the full interval of the search.  The
orientation showed no significant variations over the program, while the
pointing repeated to better than a single WFC pixel; as we discuss below, this
was actually less optimal than using a few PSFs-worth of pointing dither
between the daily visits.

\subsection{Basic Image Preparation}
\label{sec:imageprep}

The initial image reduction goal was to generate a Nyquist-sampled F814W
super-image for each WFC CCD from the dither sequence for each daily visit.
This task included repair of charge traps, hot pixels, and cosmic ray events
prior to the actual reconstruction of the super image.  Fortunately, the
dithers were executed with sufficient accuracy that this later step could be
simply done by interlacing the four images within the dither sequence.  Each
original image contributed one pixel to each a $2\times2$ pixel box in the
super-image; the scale of the super-image is twice as fine as that of the
source images.  Lauer (1999a) presents an algorithm for constructing this image
had the dithers not been exact 0.5 pixel steps, but in practice this method was
not required.

Since the images in each dither set had slightly different pointings, the
standard method of removing cosmic ray events (CRE) by comparing two exposures
with identical positioning and integration times could potentially misidentify
point sources moving with respect to the CCD pixels as cosmic ray events.  For
the WFC images in the present data set, an initial interlace image was
constructed, and the intensity of each pixel was compared to the average of its
neighbors.  Pixels that were discrepant at the $7\sigma$ level (where $\sigma$
was estimated from a WFPC2 noise-model, rather than from the deviation about
the average), {\it and} that had a value in excess of $1.6\times$ that of the
average (to avoid flagging the peaks of point sources centered on one pixel in
the sequence), were flagged as CRE.  Pixels neighboring any given hit in the
individual exposures were considered to be part of the same event if they
deviated from the interlaced image neighboring pixel average by $2.5\sigma.$
After the initial round of CRE identification, pixels affected by the hits were
deleted from the average neighboring pixel frame, and additional events were
identified in two more rounds of CRE identification.  After all CRE were
identified, affected pixels were replaced by the average of the remaining
unaffected neighbors in the interlace frame.  In practice this procedure
appeared to work extremely well for removing CRE.

In the case of the PC1 data, as the dither steps were only slightly larger than
the PC pixel scale, detection and repair of CRE were easier.  The images in a
given dither set were simply compared under the assumption that the offsets
were a single pixel in amplitude.  The average pixel values used to replace the
CRE in this regard should be better estimates than those used in the WFC dither
sets.  After the CRE were repaired, the individual PC1 images in the dither set
were shifted to a precise common origin by sinc-function interpolation (e.g.\
Castleman 1995) and combined.

During an initial reduction of the complete dataset, the brighter hot pixels
were often identified as CRE.  True hot pixels were identified as deviant
pixels that appeared at a constant CCD location over the duration of the
observations.  Unfortunately, a large population of low-level hot pixels
escaped initial detection; an additional population of hot pixels consisted of
those that newly arose within the month-long duration of the program.  Most of
these could be identified by visual inspection of the interlaced super-images.
Residual hot pixels in the interlaced image for any given day made a
readily-identifiable artifact consisting of a $2\times2$ block of elevated
sub-pixels.  When one day's interlaced image was blinked against those
surrounding it in time, the small day-to-day pointing variations additionally
helped to isolate hot pixels from compact astronomical sources.  The program
had actually specified pointing varying by $0\farcs05$ from day to day, thus
the small pointing differences were fortuitous for identifying hot pixels.  The
pointing errors were typically less than $0\farcs005$.  Random variations in
the lowest-level hot pixels, however, made them difficult to distinguish from
true point-sources with this small amount of pointing jitter; residual
hot-pixels are the most important source of false positive detections of
variable sources.  A more optimal program design would include a somewhat
larger pointing dither between repeat visits to allow for complete decoherence
between source and detector structure.

\subsection{Preparation of the Images for Variable Source Detection}

Detection of variables in M87 is discussed in detail in the next section, but
briefly variables are identified by examining the temporal run of residual
intensity values at any pixel location after the WFC interlaced images and PC1
stacked images have been registered to a common origin, have had the average
intensity value subtracted, and were processed with an optimal filter.

As noted above the pointing varied slightly from day to day.  Fortunately, the
rich M87 globular cluster system provided ample astrometric references.
Centroids of a few dozen clusters in each CCD field allowed precise angular
offsets to be derived for each day's images (presented in
Table~\ref{tab:journal}).  The roll angle was held fixed over this interval to
better than 0.05 degrees.  The images were then shifted to a common origin by
sinc-function interpolation; which does not degrade the resolution of
Nyquist-sampled images.  The full dataset (excluding the images for day-12,
which had excessive jitter) could then be stacked to make a precise average
image of M87, which in turn was subtracted from each day's data.  While in
principle this step might be omitted, in practice it greatly eases the
examination of each day's images by removing the strong intensity of the
background galaxy, fixed point sources, and the fine structure associated with
the envelope SBF pattern (essentially the Poisson noise associated with the
numbers of bright stars falling in resolution elements).

Processing each day's residual image with an optimal filter provides for the
best detection of a point source against a noisy background.  The relationship
between the filtered image, $F(x,y),$ and the initial difference image,
$D(x,y),$ is given as:
\begin{equation}
F(x,y)={D(x,y)*P^T\over \sqrt{N(x,y)*(P^TP)},}
\end{equation}
where $N(x,y)$ is a model of the expected backgrounds (e.g.\ surface brightness
+ read noise) at any pixel location, $P$ is the PSF, and $P^T$ is its
transpose, and $*$ indicates convolution (Castleman 1995).  The filtering
effectively performs an optimally-weighted integration of any point sources
present in the difference image.  The normalization converts the intensity
scale to significance in units of the locally-weighted dispersion, $\sigma$.
The filtered image $F(x,y)$ is essentially the detection significance for a
point source centered at $(x,y)$.  A point source will appear in $F(x,y)$ for
several $(x,y)$ around the true location, but the local maximum of $F$
indicates the best fit position of the source.  As for the CRE detection above,
the noise-image was based on the averaged image of M87 and the WFPC2 detector
properties.  PSFs were calculated using the TINY-TIM package; for the
subsampled WFC images, the Lauer (1999b) pixel response function was applied to
the TINY-TIM PSFs to provide the best fidelity on the diffraction scale.

\section{Selection of Candidate Events}
\label{sec:candidateselection}

The optimally filtered images described in the previous section are now
analyzed as a time series for the extraction of variable sources.  The analysis
proceeds in several steps.  First, a baseline is calculated at each pixel.
Next, pixels that exhibit consecutive significant deviations from the baseline
are recorded and grouped, and centroids are estimated.  These are the level-1
candidates.  Each level-1 candidate lightcurve is classified according to a
number of template fits.  Furthermore the level-1 candidates are compared to a
hot pixel list.  The resulting candidates are the level-2 candidates.  These
are visually inspected, eliminating obvious subtraction artifacts that are not
hot pixels.  The remaining candidates are considered real astrophysical
sources.

\subsection{Baseline Selection}
\label{sec:baseline}

The first stage in the search for variable sources is the selection of the
baseline level.  A source that flares will have several bright points, which
are included in the reference image.  Thus, the baseline flux will be lower
than that in the reference image.  We have studied several criteria for setting
the baseline, and settled on one, as we now describe.

The most naive baseline selection is obviously that the reference image is the
baseline.  This is unsatisfactory since for a given peak flux, the baseline
would depend on the timescale.

A better choice for the baseline would be to take a subset of the individual
fluxes and call the average the baseline.  For example, taking the average of
the ten lowest points on the lightcurve as the baseline works reasonably well.
As only one third of the data is involved, even slowly varying sources should
be detected.  However, this baseline selection is also unsatisfactory, as it is
quite vulnerable to downward fluctuations in flux: in fact it selects for them.

We improve the ``lowest ten'' baseline selection of the previous paragraph by
requiring that the ten points be consecutive.  We allow wraparound for
candidates that flare in the middle of the time series, e.g.\ the average of
the first four and the last six points can be the baseline.  This selection is
less vulnerable to downward fluctuations due to the consecutivity.  As such, we
use this definition of the baseline level in all of the following analysis.
Varying the number of samples taken doesn't have much effect, though taking too
many eliminates slowly varying sources and taking too few makes the baseline
quite noisy.  Essentially, we are constructing a running reference image of ten
individual images, and taking the lowest such image for each pixel individually
as the baseline.

\subsection{Variability Selection}
\label{sec:variabilityselection}

Having set the baseline using the running reference image as in
\S\ref{sec:baseline}, we now search for sources that vary about the baseline
significantly.  We study two minimum thresholds of $\Delta\chi^2=50$ and
$\Delta\chi^2=100$ relative to a baseline only fit, or equivalently a
signal-to-noise ratio $Q$ of about 7 and 10, respectively.  The required
signal-to-noise ratio can be accumulated over several images, and we use
several different tests.

We first apply a basic consecutivity test, namely several consecutive images
exhibiting a certain significance of detection so that the total gives the
threshold $\Delta\chi^2$.  By requiring only a single point, we find that there
is too much sensitivity to hot pixels, artifacts, cosmic rays and the like.
Requiring two consecutive detections of $Q=\sqrt{\Delta\chi^2/2}$ gives the
same total significance and rejects many false detections while allowing fast
events to be detected.  This is our basic criterion for variability.

We modify this consecutivity test to allow longer timescale events that might
be dimmer, though with similar total significance.  We test for five
consecutive $Q=\sqrt{\Delta\chi^2/5}$ detections, and find more candidates.
Extending this test to eight consecutive samples does not yield any new
candidates.

These consecutivity tests are sensitive to downward fluctuations ending the
consecutive streak.  As a check, we use one final test in which the images are
averaged in consecutive groups of three, yielding three series of images (1-3,
4-6 etc., 2-4, 5-7, etc., and 3-5, 6-8, etc.)  In terms of signal-to-noise
ratio, the new image is $Q_{\rm 1-3}=(Q_1+Q_2+Q_3)/\sqrt{3}$.  The two
consecutive $Q=\sqrt{\Delta\chi^2/2}$ test is then applied to these averaged
image series.

We have experimented with other similar criteria, e.g.\ four consecutive, and
also requiring a peak sample at higher significance, e.g.\ five consecutive
$3\sigma$, including at least one $5\sigma$.  No new candidates are found in
the tests we performed.

We denote sources identified by one or more of the consecutivity tests as the
level-1 candidates.  There is a high level of duplication here, as we have not
grouped candidate pixels together at this stage.  This can be done simply by
sorting pixels by their $x$ and $y$ coordinates.  An approximate centroid is
also calculated at this stage.

\subsection{Lightcurve Fitting}
\label{sec:lightcurvefitting}

Once the level-1 candidates are identified, they are classified according to
several template lightcurves.  We use four basic templates.  First is the
trivial constant baseline.  The free parameter is simply the baseline level,
and the best fit is the average of the points.  A two-level baseline (step
function) is good for flagging hot pixels.  The free parameters are the two
baseline levels and the time of the step.  We use a linear ramp, though very
few level-1 candidates have this as the best fit.  There are two free
parameters, the slope and intercept.  Lastly, we fit the degenerate
microlensing lightcurve (which takes the peak magnification to infinity holding
the peak flux constant).  The parameters are the baseline, the peak time, the
peak flux, and the full-width at half maximum timescale.

Each level-1 candidate is fitted to each template, with $\chi^2$ being
calculated.  Most level-1 candidates have either the two-level baseline or the
microlensing template as the best fit.  We discard candidates whose best fit is
not the microlensing template, and furthermore we discard events where
$\chi^2({\rm other})-\chi^2({\rm microlens})<0.25/\rm dof$.

\subsection{Hot Pixels}
\label{sec:hotpixeltest}

The WFPC2 has a sizable number of hot pixels, which may be confused with
astrophysical variable sources.  Bright hot pixels are removed along with
cosmic ray events (described in \S\ref{sec:imageprep}).  It is the dimmer hot
pixels that are troublesome.  Many of these are active at a very low level, and
not much concern for imaging, though of crucial concern for a variability
search.  We attempt to compare sources in the difference images with the PSF to
separate the hot pixels from real sources.

We use the difference images directly to identify hot pixels.  As many of them
are not very active, we average the difference images in consecutive groups of
five to improve the sensitivity.  If a hot pixel is discovered in any of these
stacks, it is flagged so that candidate events at its position can be
discarded.

We use a PSF test to identify hot pixels.  In the dithered images, a hot pixel
should appear as a $2\times2$ box.  A star (PSF) is more extended than this.
We calculate the ratio of the average pixel values in the central four pixels
to the average pixel values in the surrounding twelve pixels.  In principle
this is infinite for a hot pixel.  A PSF yields a finite value of roughly 2.5
(the central four pixels are on average 2.5 times brighter than the surrounding
twelve).  Allowing for Poisson fluctuations, we flag pixels as hot if their
center-surround ratio is more than $5\sigma$ above the expected value for a
PSF.  Furthermore, we require that hot pixel is symmetric (again allowing for
Poisson fluctuations): this significantly reduces the incidence of bright
variable sources being flagged as hot pixels on their outskirts.

\subsection{Visual Inspection}

The level-1 candidates that have a best fit microlensing lightcurve with an
acceptable $\chi^2$, and are not flagged as hot pixels are denoted the level-2
candidates.  These candidates (there are only a small number) are then visually
inspected.  The majority are found at the edges of globular clusters in the
images.  This is a known difficulty.  In regions of high brightness gradient,
the noise level is underestimated because the PSF$^2$ is much more sharply
peaked than the PSF, which thus samples the bright center.  The subtle PSF
variations from visit to visit are then estimated to be of higher statistical
significance than they should be, occasionally producing false positives.  The
globular clusters are only barely resolved.  In principle we could have marked
the globular clusters as ``hot pixels'' to alleviate this, but the number of
such candidates is small.

At this stage, most of the remaining candidates can be visually identified as
hot pixels, as they exhibit a $2\times2$ pixel pattern that is fixed in CCD
coordinates.  These necessarily had statistical fluctuations that allowed them
to pass the crude hot pixel test.

To identify true variable sources, we have used the fact that the images in the
stack are misaligned slightly from visit to visit.  Thus true variable sources
remain fixed in the frame of the globular clusters (which is moving relative to
the frame of the CCD).

\subsection{Candidate Events}

After all tests have been applied, seven candidate astrophysical sources
emerge.  One has an excellent microlensing fit, two appear to be novae, and
four sources have rising or declining lightcurves over the 30 days, and might
be novae or perhaps variable stars.

\begin{figure}
\epsfig{width=0.95\textwidth,file=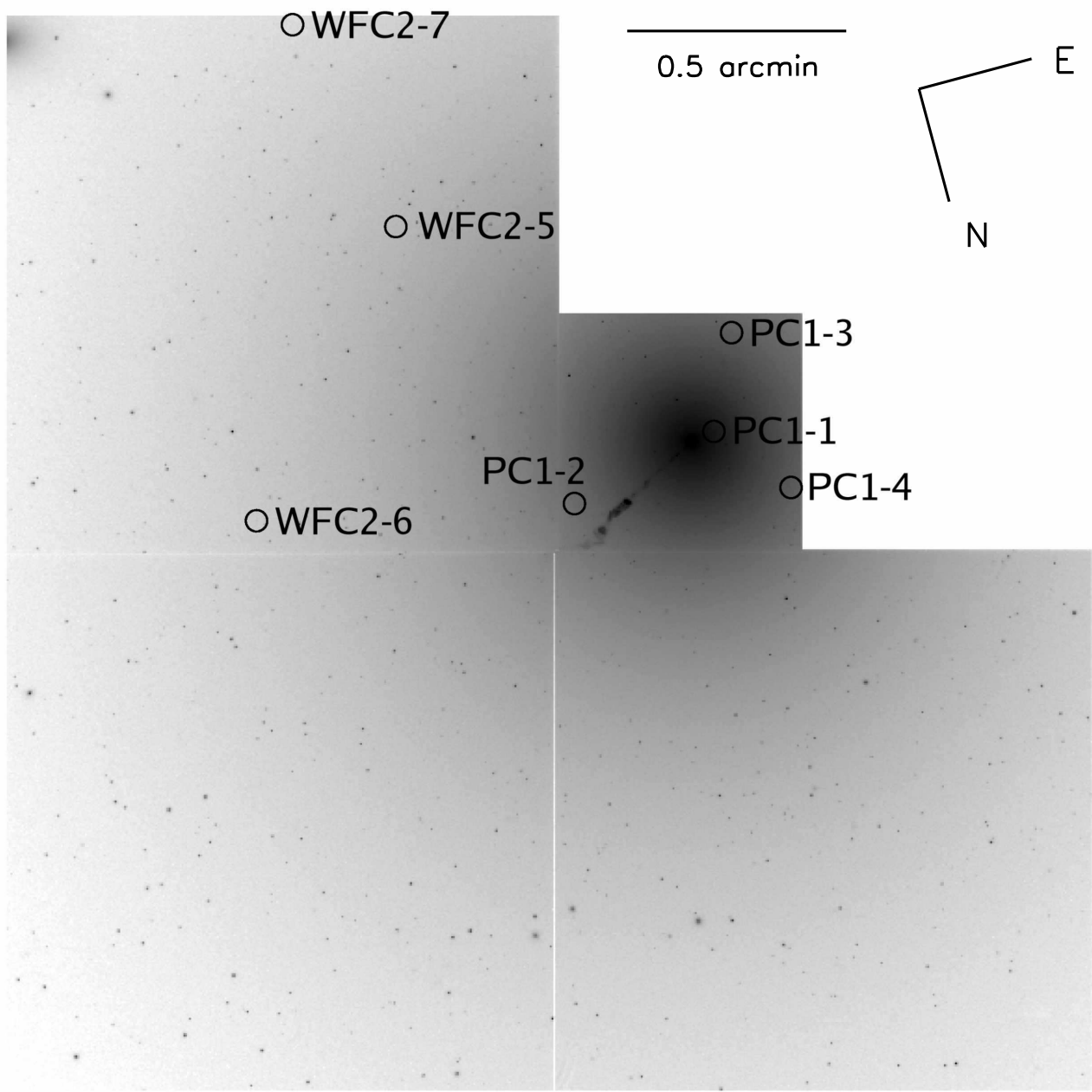}

\caption{Finder chart for the seven candidates in PC1 and WFC2 chips.
}
\label{fig:findercharts}
\end{figure}

\begin{figure}
\begin{center}\epsfig{width=0.95\textwidth,file=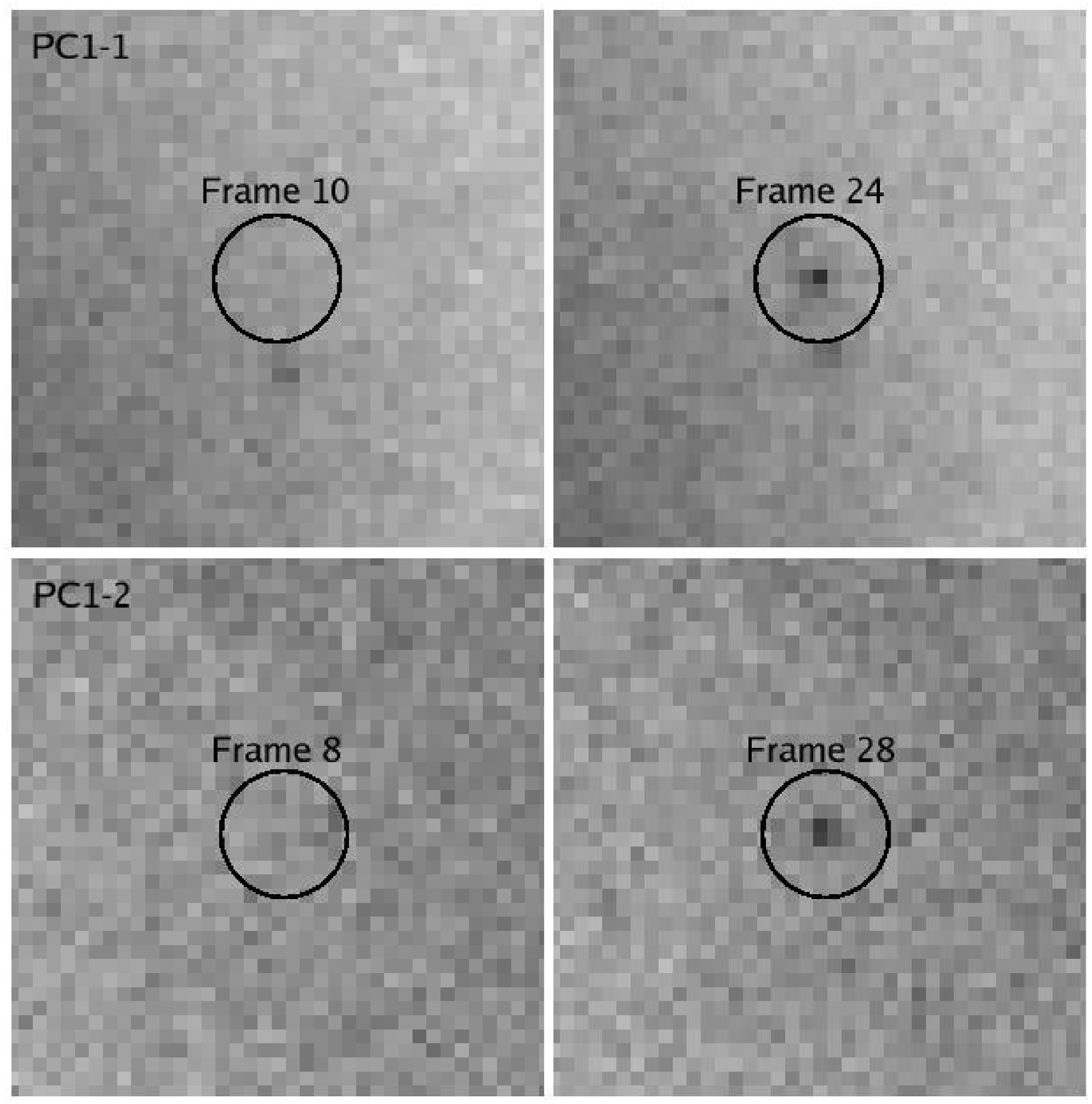}\end{center}

\caption{Unsubtracted images for events PC1-1 and PC1-2.}
\label{fig:ev_12}
\end{figure}

\begin{figure}
\begin{center}\epsfig{width=0.95\textwidth,file=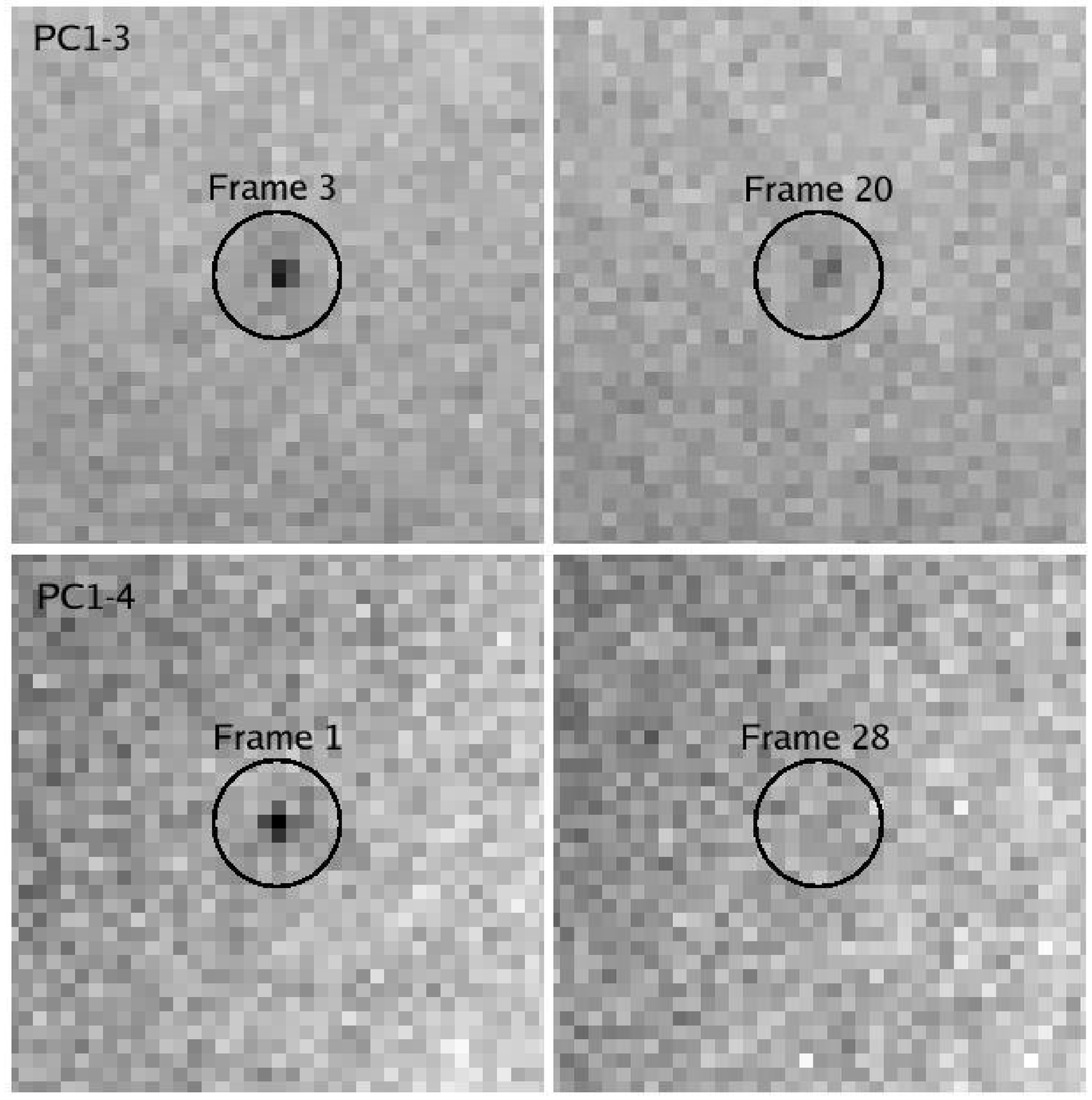}\end{center}

\caption{Unsubtracted images for events PC1-3 and PC1-4.}
\label{fig:ev_34}
\end{figure}

\begin{figure}
\begin{center}\epsfig{width=0.95\textwidth,file=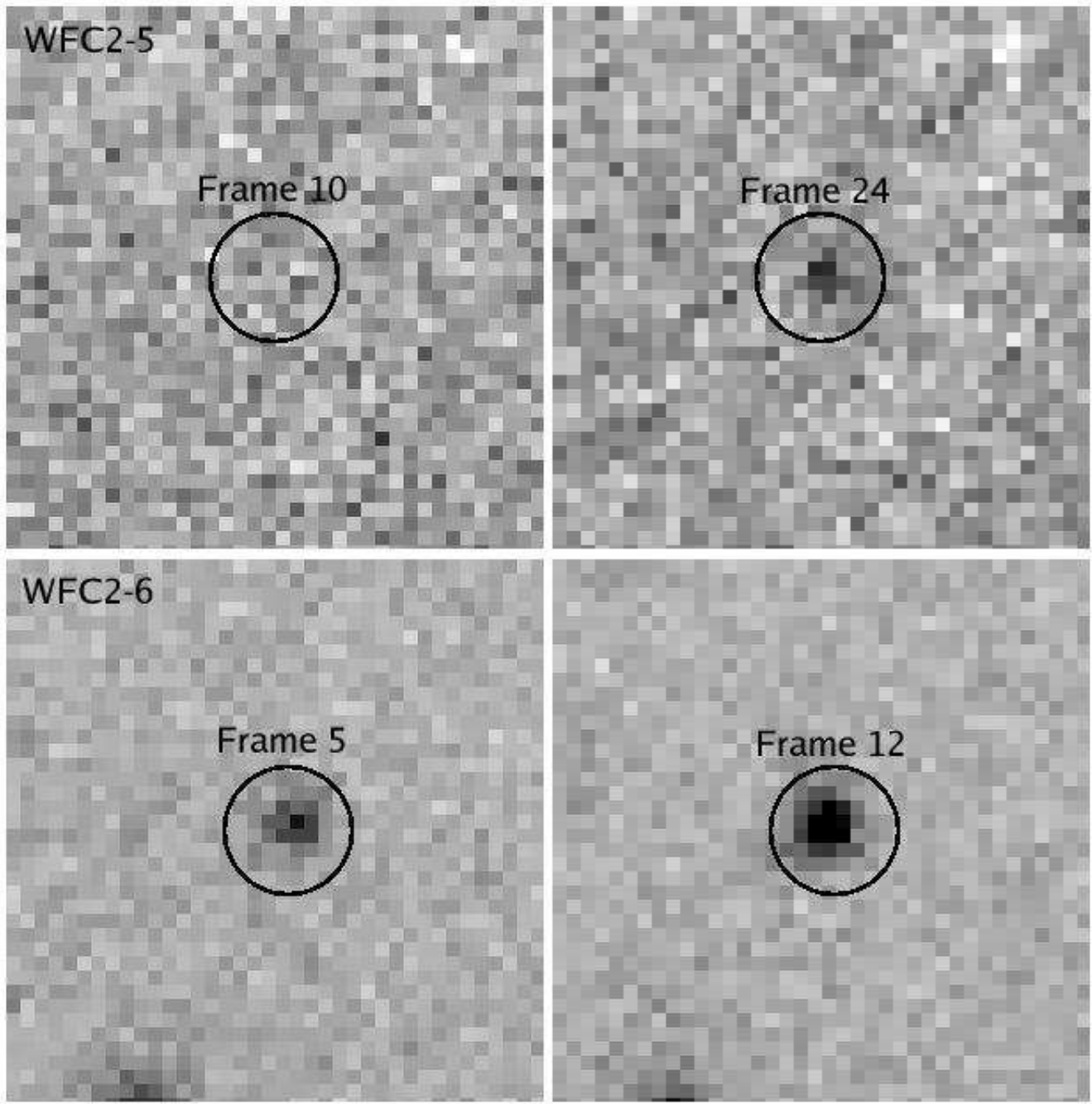}\end{center}

\caption{Unsubtracted images for events WFC2-5 and WFC2-6.}
\label{fig:ev_56}
\end{figure}

\begin{figure}
\begin{center}\epsfig{width=0.95\textwidth,file=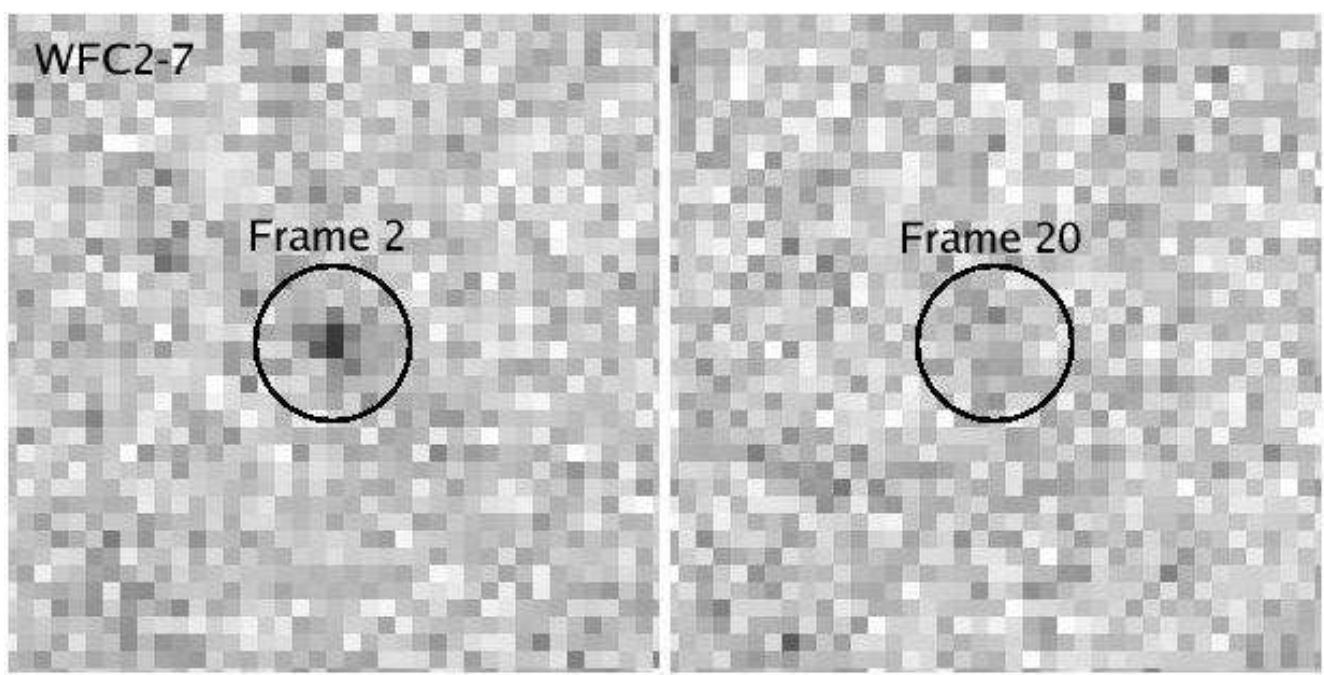}\end{center}

\caption{Unsubtracted images for event WFC2-7.}
\label{fig:ev_7}
\end{figure}

With a threshold $\Delta\chi^2=50$, even with the hot pixel test applied, there
are a number of ambiguous candidates.  Among these there are the seven
candidates that are clearly astrophysical.  Increasing the threshold to
$\Delta\chi^2=100$ removes most ambiguous detections, but allows all seven
clear candidates.  Thus, we will err on the side of conservatism and take the
threshold to be $\Delta\chi^2=100$.  The candidates are listed in
Table~\ref{tab:candidate}, along with their microlensing fit parameters in
Table~\ref{tab:candidatefits}.  A mosaic finder chart for the seven candidates
is presented in Fig.~\ref{fig:findercharts}.  In
\mbox{Figs.~\ref{fig:ev_12}-\ref{fig:ev_7}} we illustrate the unsubtracted
frames for the seven candidate events for both the baseline and the peak flux.
Note that PC1-3 is a resolved source, and is visible in each frame before
subtraction, and that WFC2-6 is coincident with a globular cluster.

\subsection{First Interpretation of Candidates}

Our primary science goal is to study microlensing populations around M87.  With
this aim, it is now appropriate to reject the candidates whose microlensing
fits have low probability according to the $\chi^2$ distribution.  Rejected
events remain interesting as nova or variable candidates, as the microlensing
template is a good generic bump finder.  Non-microlensing bumps will pass all
tests except that their microlensing fits will be unlikely.  We require a
$\chi^2$ within the usual $2\sigma$ confidence.  For the 26 degrees of freedom
appropriate for the microlensing fit, the requirement is $0.5253<\chi^2/{\rm
dof}<1.6277$.  In fact the application of the lower limit does not exclude any
events.

These variable sources have been selected in the F814W frames.  They are now
sought in the F606W frames, and approximate $V-I$ colors determined by simple
aperture photometry.  We can detect sources in F606W at the same or greater
significance as F814W if $V-I\approx0.7$ or bluer.

\begin{figure}
\epsfig{width=0.49\textwidth,file=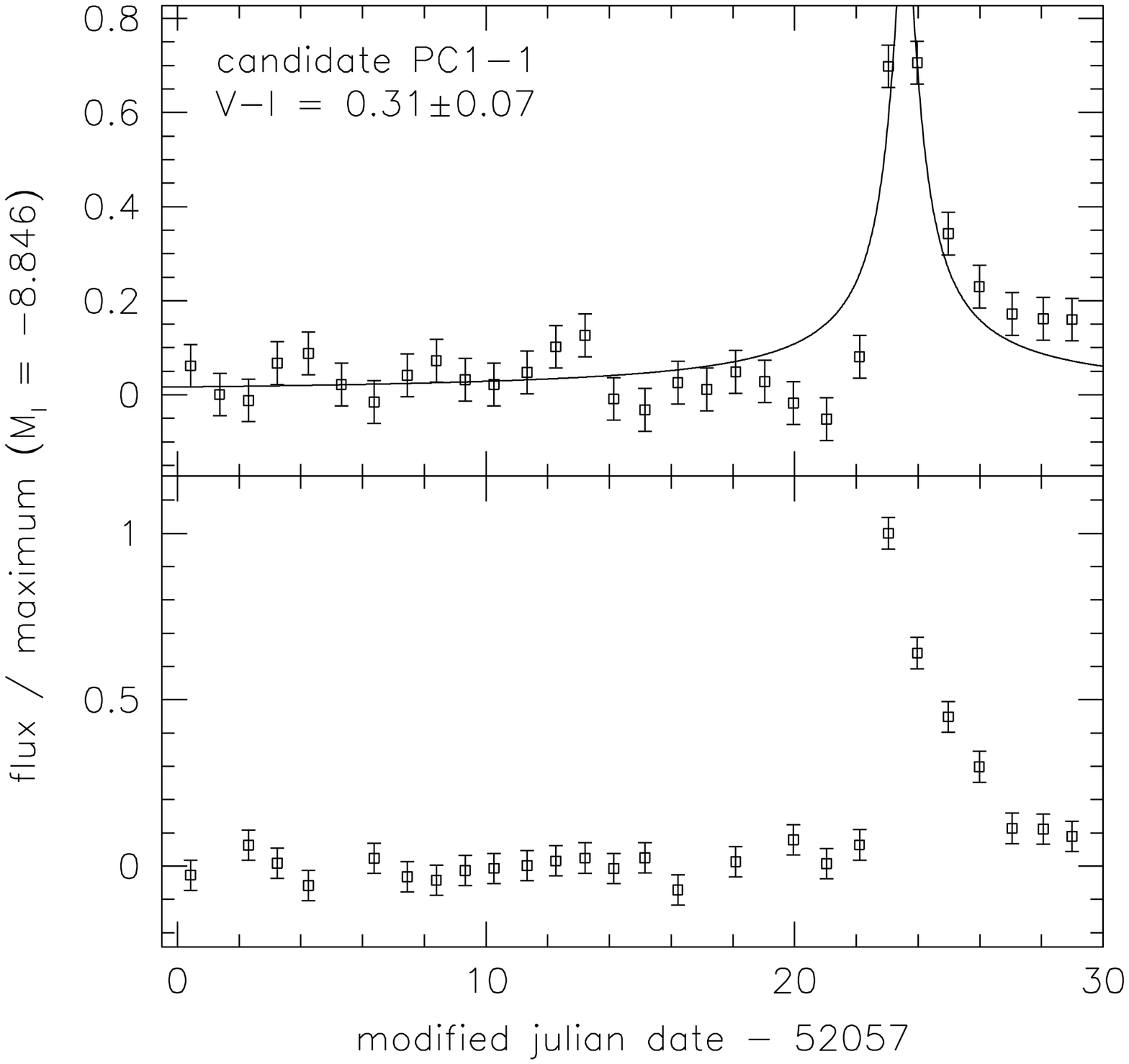}
\epsfig{width=0.49\textwidth,file=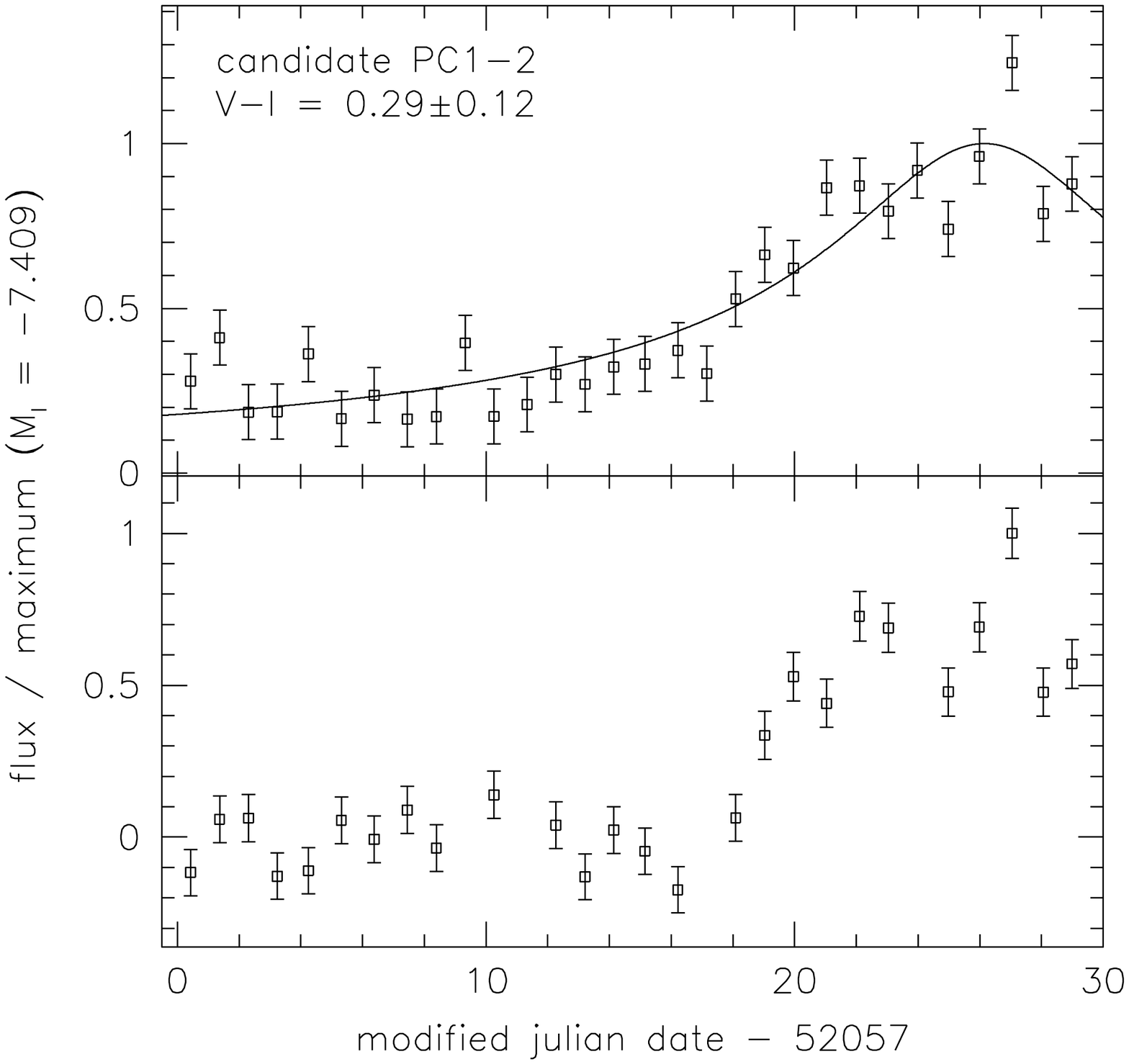}
\epsfig{width=0.49\textwidth,file=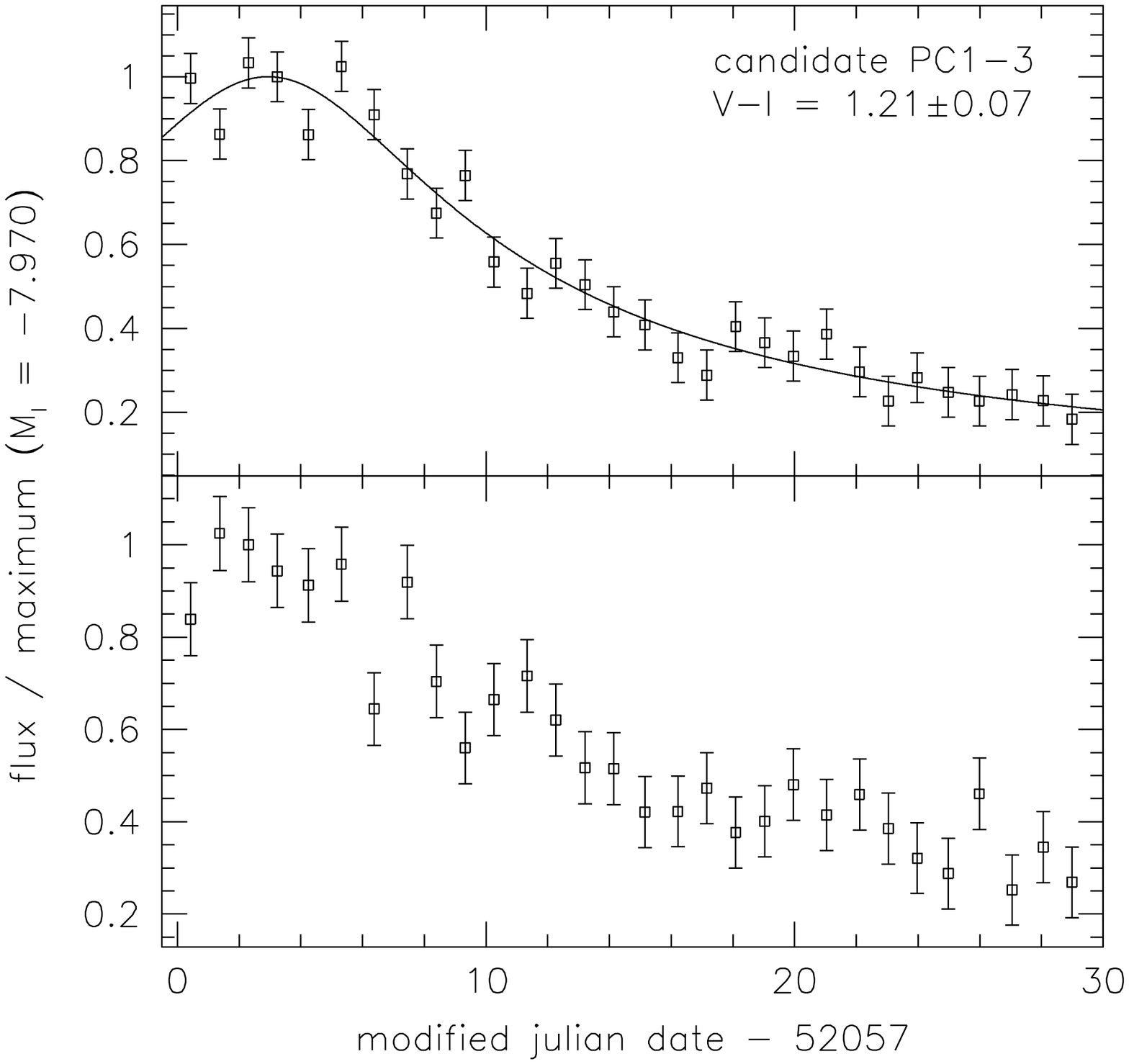}
\epsfig{width=0.49\textwidth,file=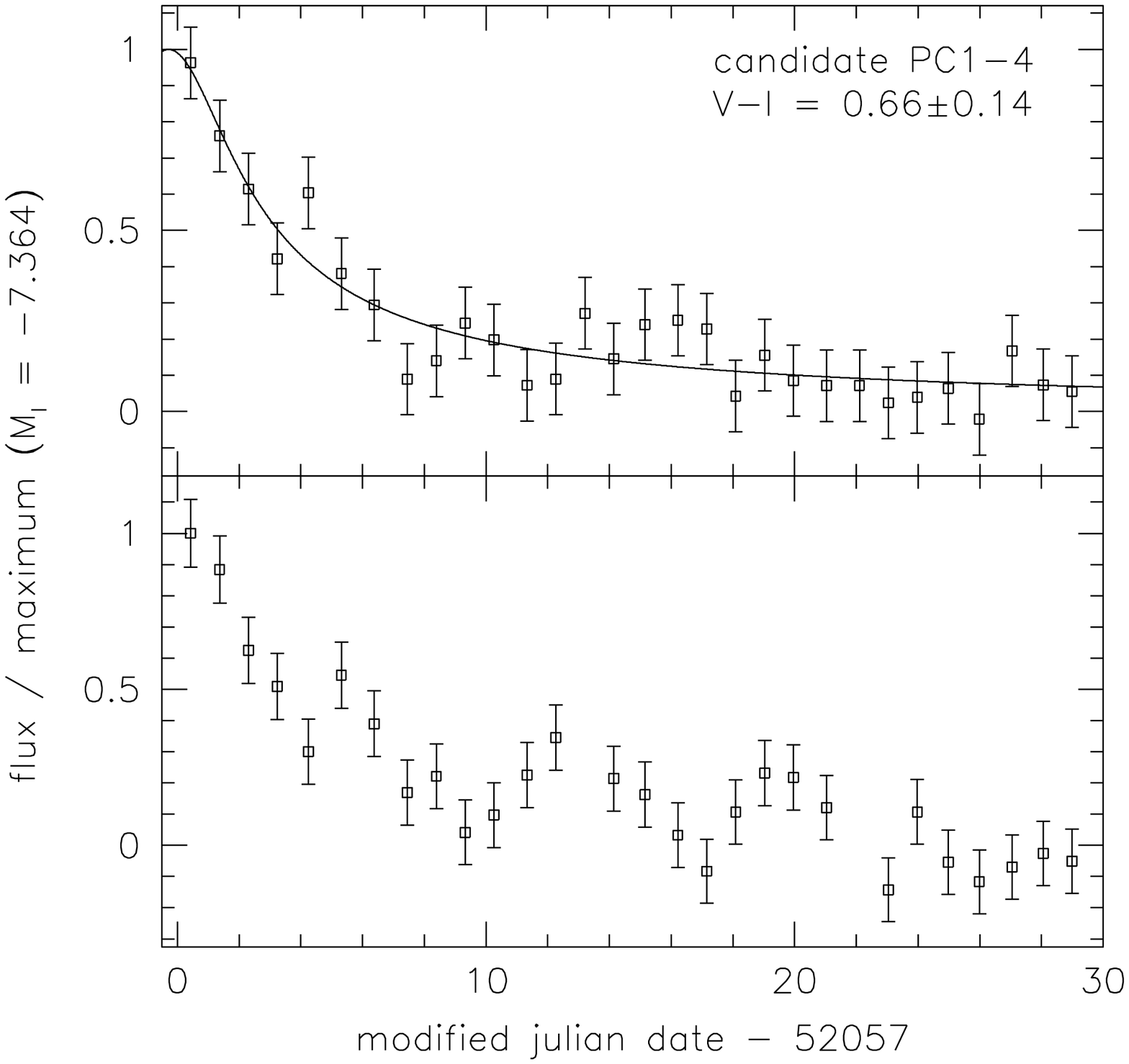}

\caption{Candidates in PC1.  The top panels illustrate the F814W data, along
with the microlensing fits (solid curve).  The bottom panels illustrate the
F606W data.}
\label{fig:pc1}
\end{figure}

\begin{figure}
\epsfig{width=0.49\textwidth,file=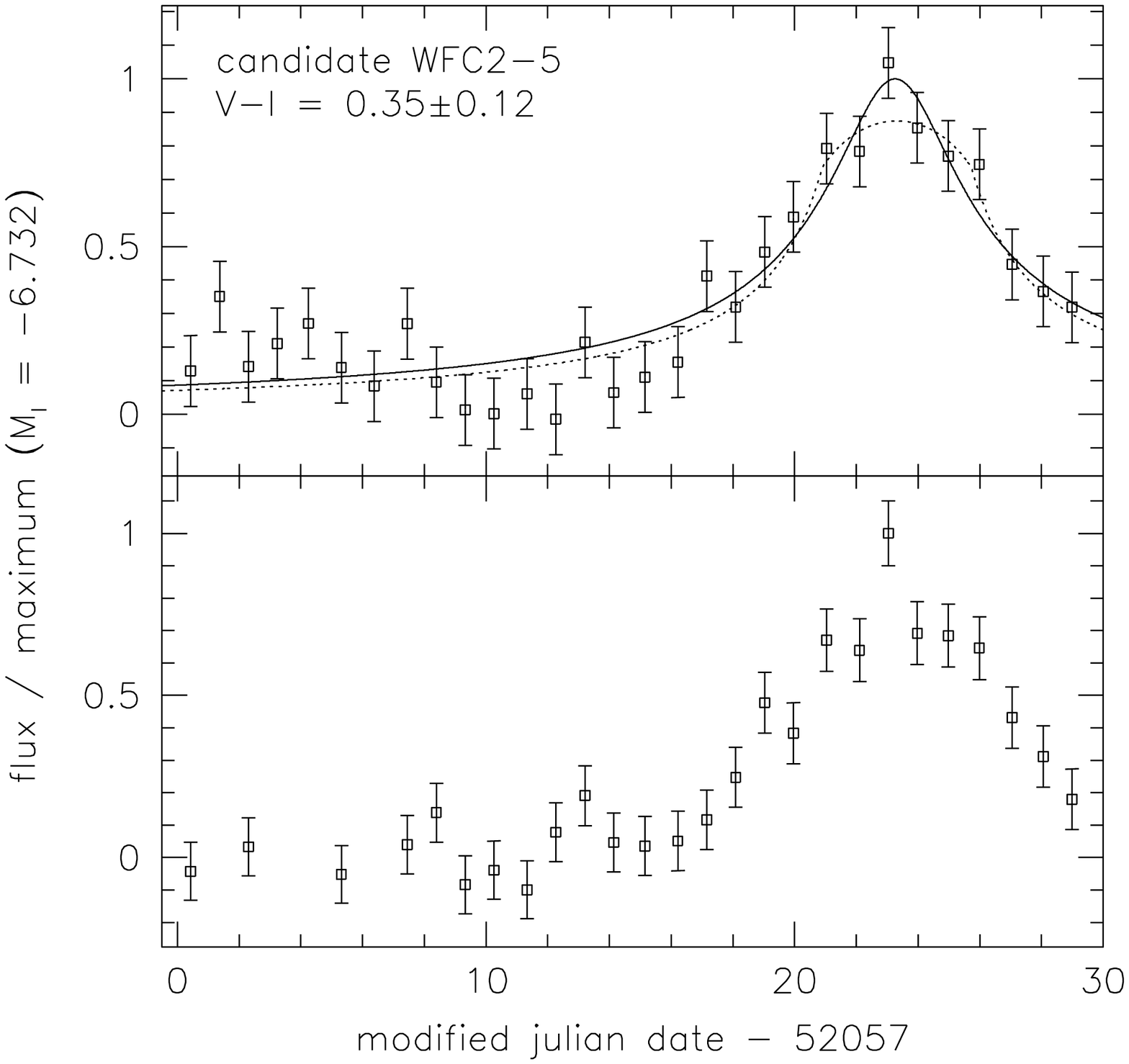}
\epsfig{width=0.49\textwidth,file=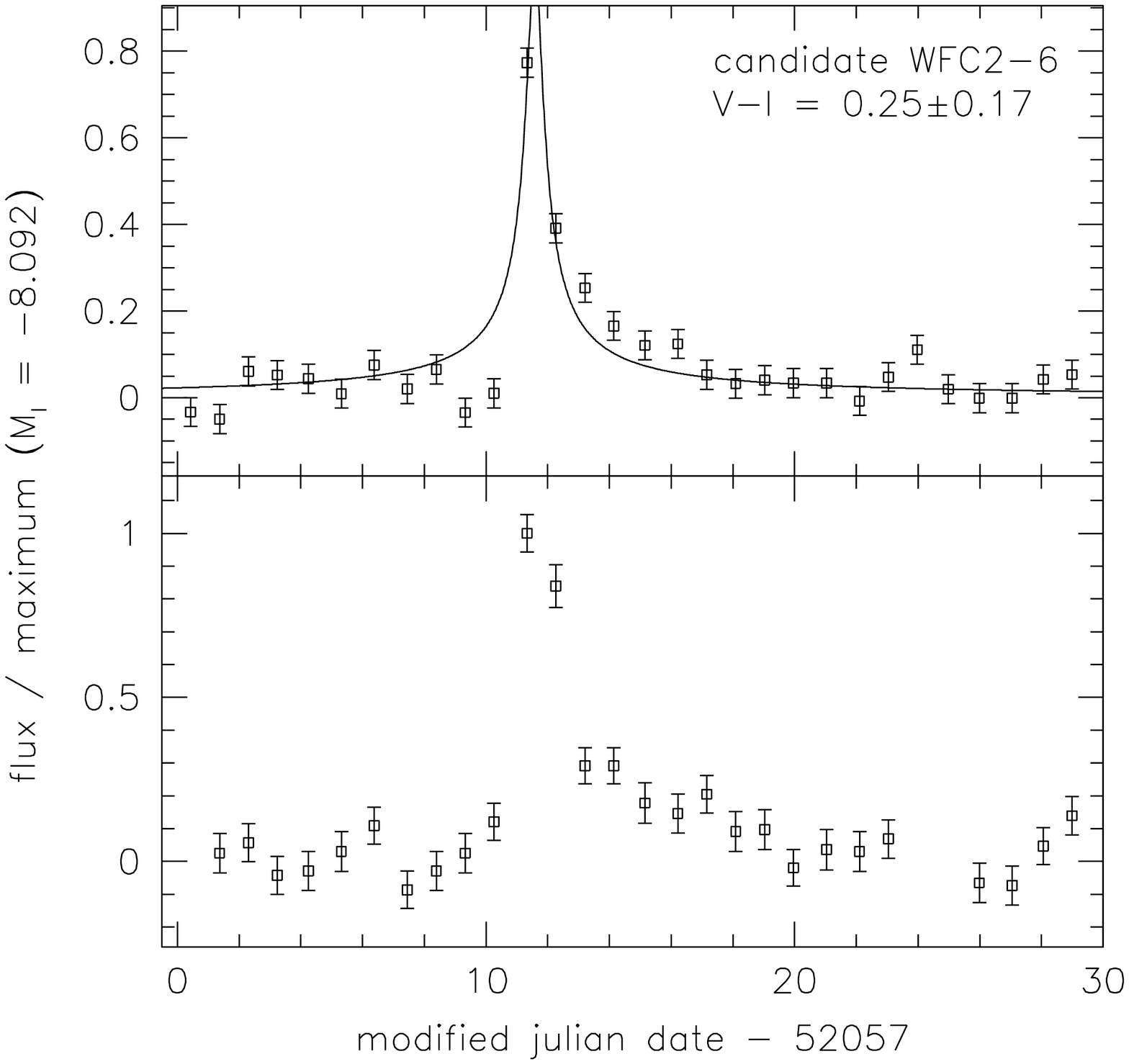}
\centerline{\epsfig{width=0.49\textwidth,file=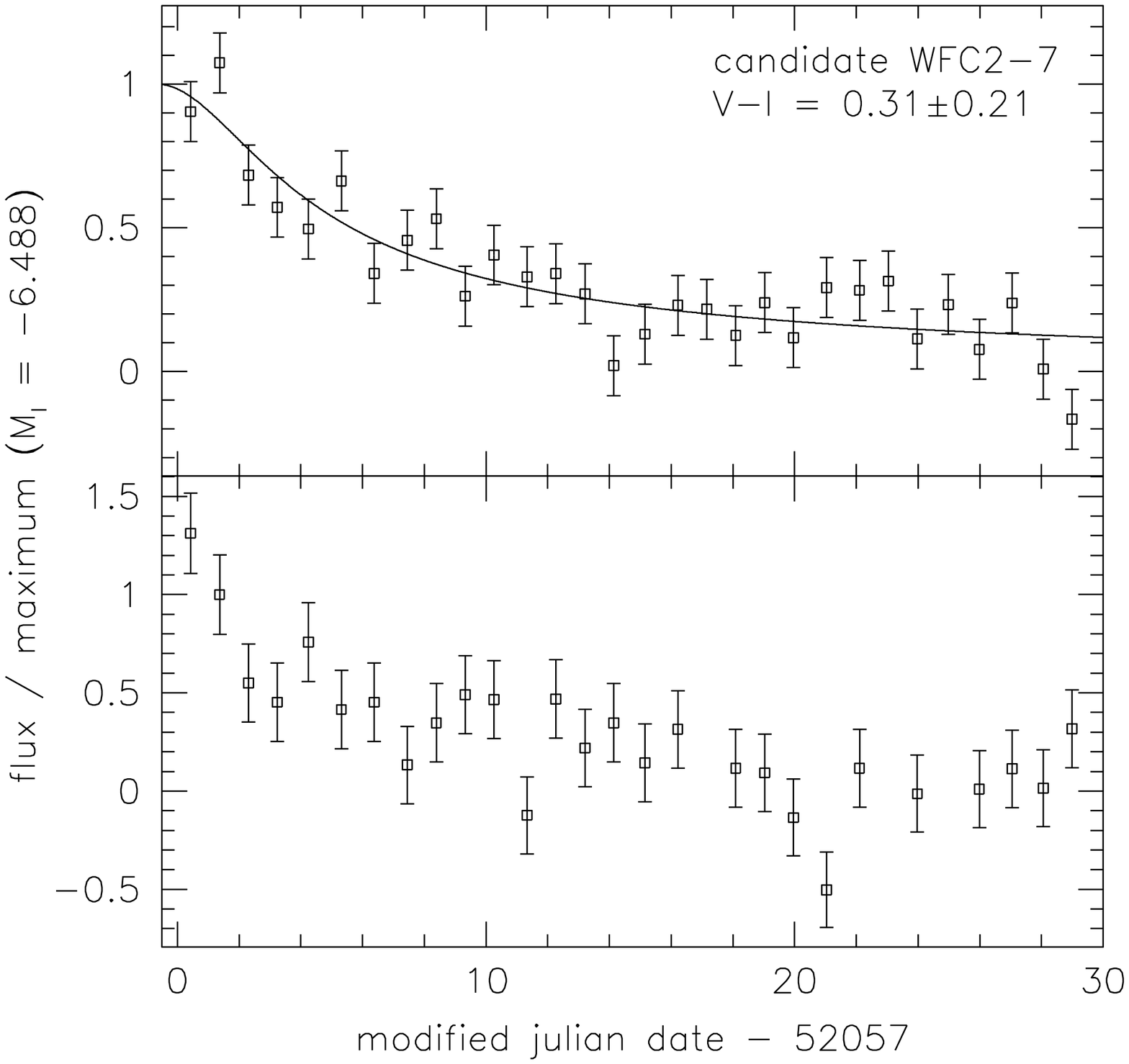}}

\caption{Candidates in WFC2.  The layout is the same as Fig.~\ref{fig:pc1}.
For the microlensing candidate WFC2-5, the finite source fit is given (dotted
curve).}
\label{fig:wfc2}
\end{figure}

\begin{figure}
\epsfig{width=0.49\textwidth,file=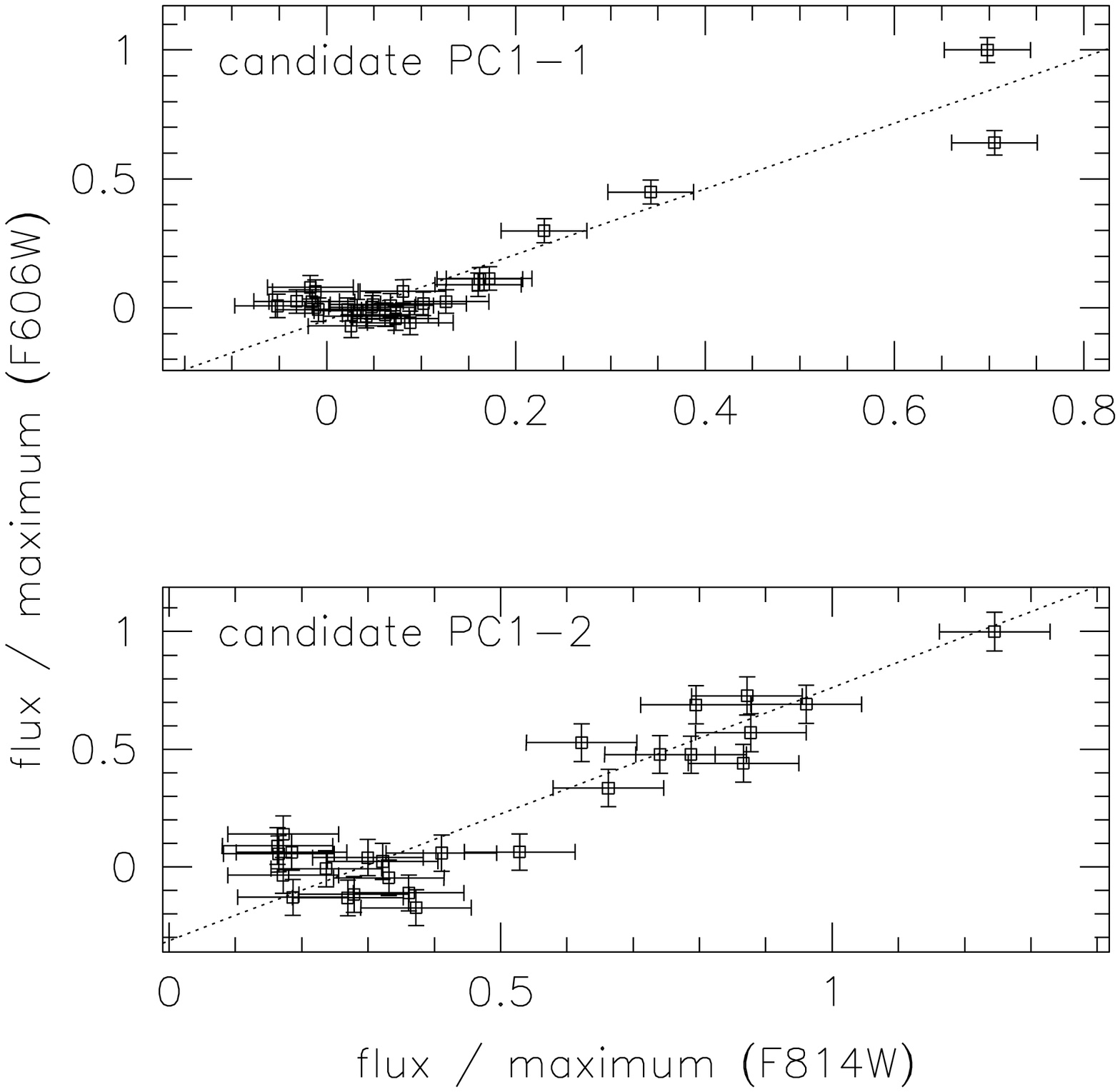}
\epsfig{width=0.49\textwidth,file=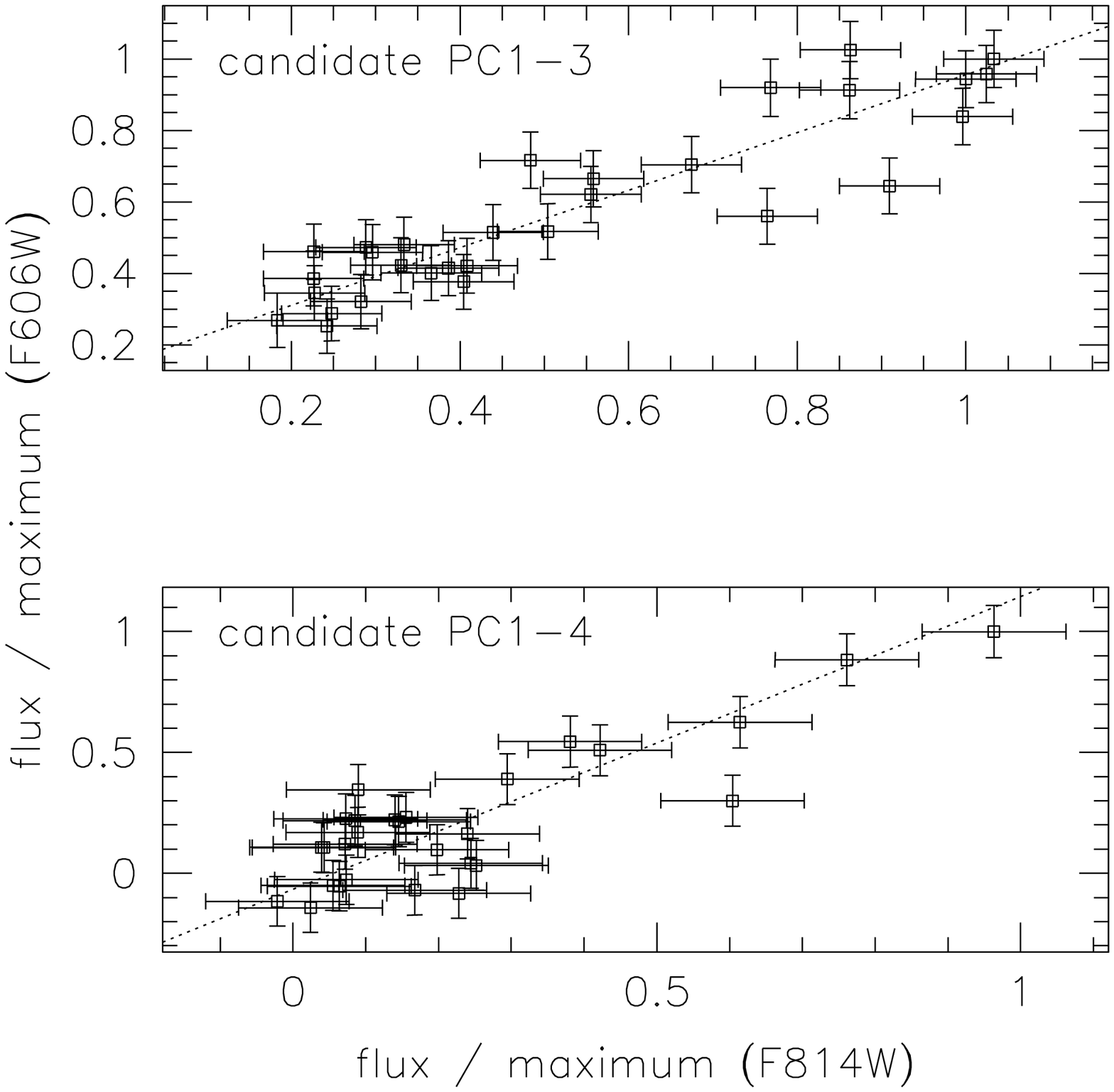}
\epsfig{width=0.49\textwidth,file=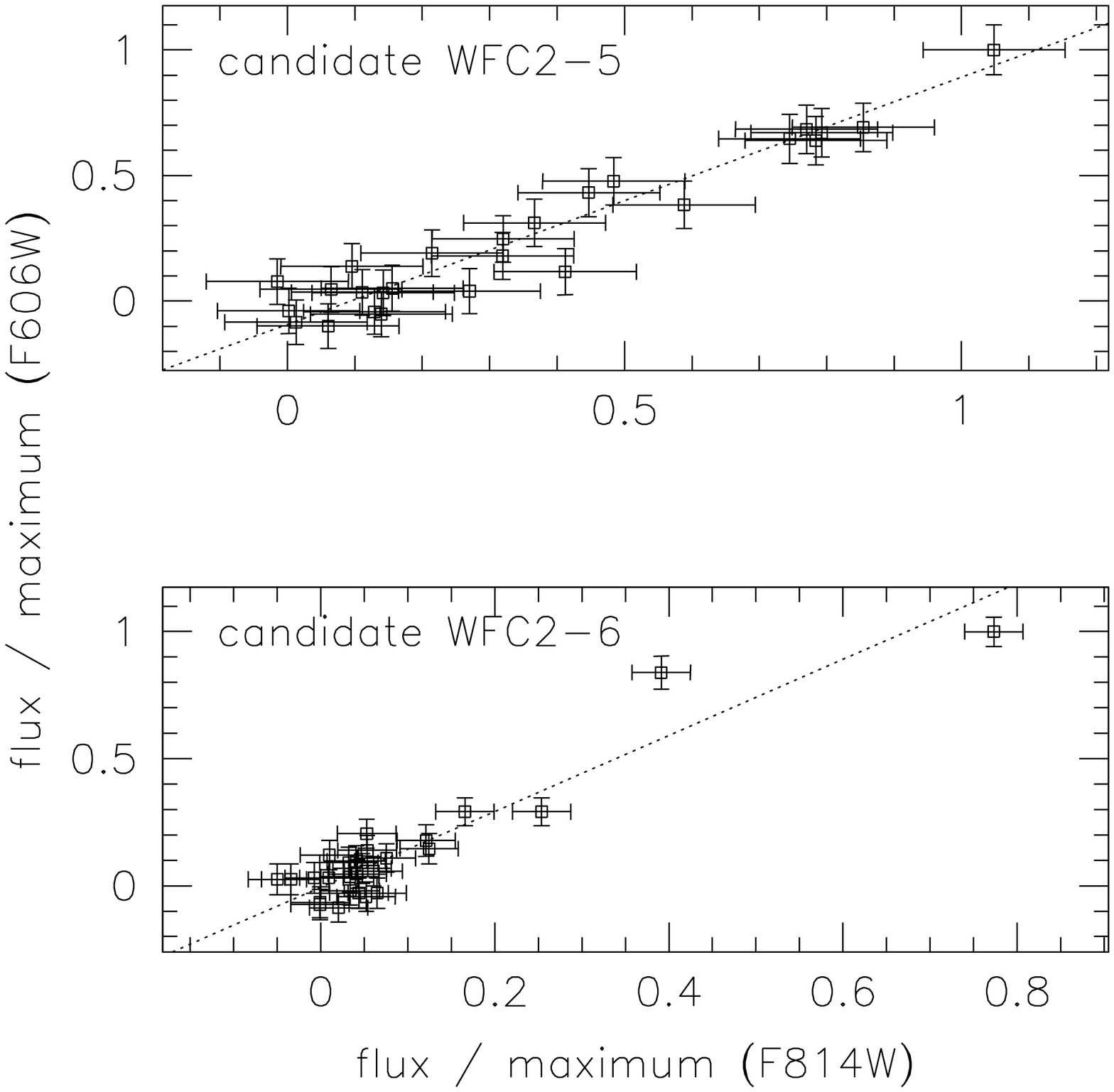}
\epsfig{width=0.49\textwidth,file=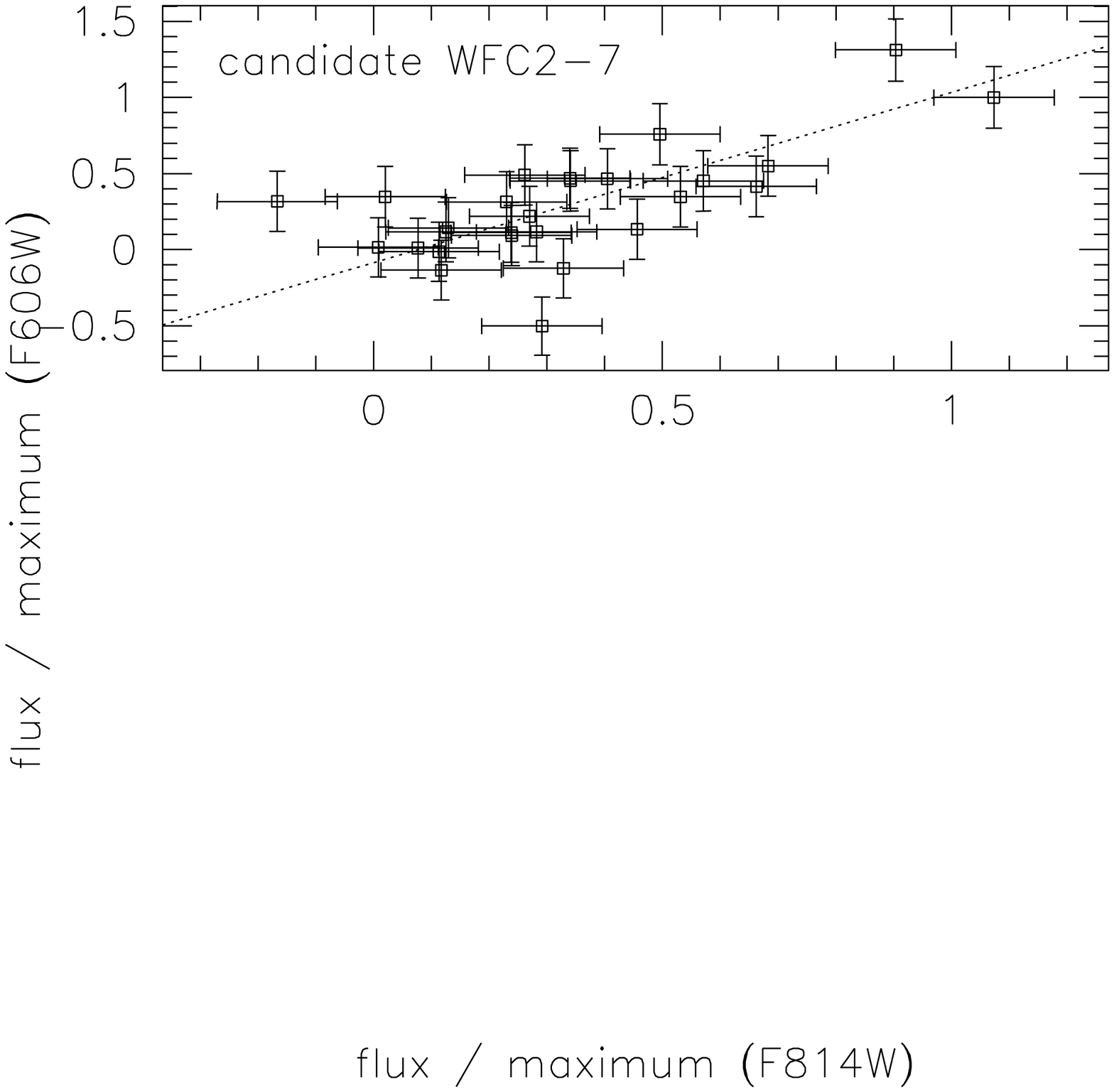}

\caption{F606W flux vs.\ F814W flux for the candidate events, with best linear
fit (dotted line).  A microlensing event should exhibit a straight line.  Any
achromaticity would give a deviation from a linear relation.  Only the two
clear nova candidates (PC1-1 and WFC2-6) exhibit a clear color change.}
\label{fig:flux}
\end{figure}

The photometric fit parameters are listed in Table~\ref{tab:candidatefits}.
Note that the flux {\em excess} is listed: this is the flux above the baseline
of the microlensing fit.  This excess flux is converted to magnitudes in the
$I$-band.  Likewise, the $V-I$ color listed is the color of the excess flux.
The lightcurves of the candidate events are plotted in
Figs.~\ref{fig:pc1}-\ref{fig:wfc2}.  The fluxes in the two filters are plotted
against each other in Fig.~\ref{fig:flux}.  Any microlensing event should
exhibit a straight line on a flux-flux plot.  Only candidates PC1-1 and WFC2-6
exhibit a significant color change.  As we discuss next, these are likely to be
novae.

Candidates PC1-1 and WFC2-6 are quite clearly novae, according to their
magnitudes, colors and lightcurve shapes.  Candidate WFC2-6 appears to be in a
globular cluster of M87, and is discussed in more detail elsewhere (Shara et
al.~2003).  Candidates PC1-2 and WFC2-7 seem to be blue variables, possibly
novae.  Since their peaks are unobserved, it is hard to say more about them.
Candidates PC1-3 and PC1-4 are redder variables, again with peak brightness
unobserved.  Candidate PC1-3 is in fact detected in all 30 visits.  Candidate
WFC2-5 is an excellent microlensing candidate, though its blue color is
unexpected: red giants with $V-I > 1$ are typically the most numerous
sources.

Candidates PC1-3, PC1-4, WFC2-5, and WFC2-7 have an acceptable $\chi^2$ to be
microlensing.  However, only candidate WFC2-5 is sampled on both sides of the
peak.  Thus, while any of candidates PC1-3, PC1-4, WFC2-5, and WFC2-7 could be
microlensing, only candidate WFC2-5 can be confidently proposed as a
microlensing event.  As a final criterion, we require that at least half of the
half-width at half maximum $(=0.25\,\thalf)$ be sampled on either side of the
peak, thus only candidate WFC2-5 remains.

\section{Detection Efficiency}
\label{sec:detectionefficiency}

Having developed the procedure for finding microlensing events in the dataset,
we must now calculate the detection efficiency.  We proceed in two ways.
Starting at the level of event lightcurves, we thoroughly model the detection
efficiency simply by generating a large number of artificial lightcurves with
known event parameters and applying the lightcurve analysis, as described in
\S\ref{sec:lightcurvetests}.  At the image level, artificial events are
generated and put into the image stack.  In this way the efficiency of the
steps between the difference images and extracting lightcurves can be
estimated, as described in \S\ref{sec:artificialstars}.  In the end, we will
use the lightcurve efficiencies, with a correction derived by comparing with
the artificial star efficiencies.

\subsection{Lightcurve Tests}
\label{sec:lightcurvetests}

Calculating the detection efficiency using only a lightcurve test necessarily
assumes that the noise model is perfect.  We proceed with this assumption, but
we will test it using artificial stars in \S\ref{sec:artificialstars}.

A large number of theoretical lightcurves are generated, taking a grid over the
interesting fit parameters: the peak significance
$\qmax=\sqrt{\Delta\chi^2\;{\rm (peak)}}$, the timescale $\thalf$, and the
minimum impact parameter $\beta$ (which has only a small effect on the shape of
the lightcurve).  Fixing these three parameters, lightcurves are generated with
random values of the peak time $t_0$, ranging over a generous interval $[t_{\rm
min},t_{\rm max}]$ containing the observation epochs.  The fluxes at each epoch
are taken from a Poisson distribution.  These artificial lightcurves are then
passed through the stages of baseline selection (\S\ref{sec:baseline}),
variability selection (\S\ref{sec:variabilityselection}) and lightcurve fitting
(\S\ref{sec:lightcurvefitting}), identically to the lightcurves produced at
each pixel in the result images.  The fraction of artificial lightcurves, all
representing ``true'' microlensing events, that pass all of these tests is then
the detection efficiency for events with $t_{\rm min}\le t_0\le t_{\rm max}$,
and will be denoted $P\left(\thalf,\qmax,\beta\right)$.  Results for some
interesting fit parameters are plotted in Fig.~\ref{fig:efficiencies}.

\subsection{Artificial Star Tests}
\label{sec:artificialstars}

As a check on the lightcurve detection efficiency of
\S\ref{sec:lightcurvetests}, we use an artificial star test.  We randomly
generate microlensing events and insert them in the existing image stack.  This
does in principle introduce a bias toward the large area of low surface
brightness, but the event rate is expected to be only a weak function of
surface brightness, so this prescription is adequate for our purposes.  This
artificial image stack is then processed identically to the true image stack.
Some fraction of the artificial events are recovered.  This fraction is the
artificial star efficiency.

We focus on the WFC2 chip.  Starting from just the artificial reference and
difference images, the simplest test to be applied is the hot pixel test of
\S\ref{sec:hotpixeltest}.  The probability that an artificial event is
identified as a hot pixel can be determined.  This analysis in fact provided
guidance on how to construct the test so that there was a low probability of a
false positive.  We use three timescales: $\thalf=5,10,15$ days, and eight peak
flux levels corresponding to $\Delta\chi^2=12.5,25,50,100,200,400,800,1600$.
For each combination, two thousand artificial events are generated.  The events
are randomly distributed uniformly in position on the chip and in peak time
(between frame 1 and frame 30).  Furthermore, they are uniformly distributed in
sub-pixel offsets in units of 0.005\arcsec along each axis.

The expected flux of the artificial event is now determined from the
theoretical lightcurve, which is in units of statistical significance.  This is
converted to expected flux simply by consulting the reference image to
calculate the noise level.  With the expected flux in each frame, multiplied by
the normalized PSF, appropriately shifted, we compute the expected number of
photons in each pixel due to the artificial event.  The actual number for each
pixel is generated according to a Poisson distribution.  In this way for each
artificial event, in each of the 30 frames, in each PSF pixel, we compute the
number of photons due to the event.  At each pixel position, we take the
average over frames (neglecting frame 12) of the artificial event photons, and
add this number to the reference image.  This accounts for the fact that all
events appear in the reference image at some level.  We set a rough baseline
for each event, and add or subtract the photons from the difference images.

\begin{figure}
\epsfig{width=0.49\textwidth,file=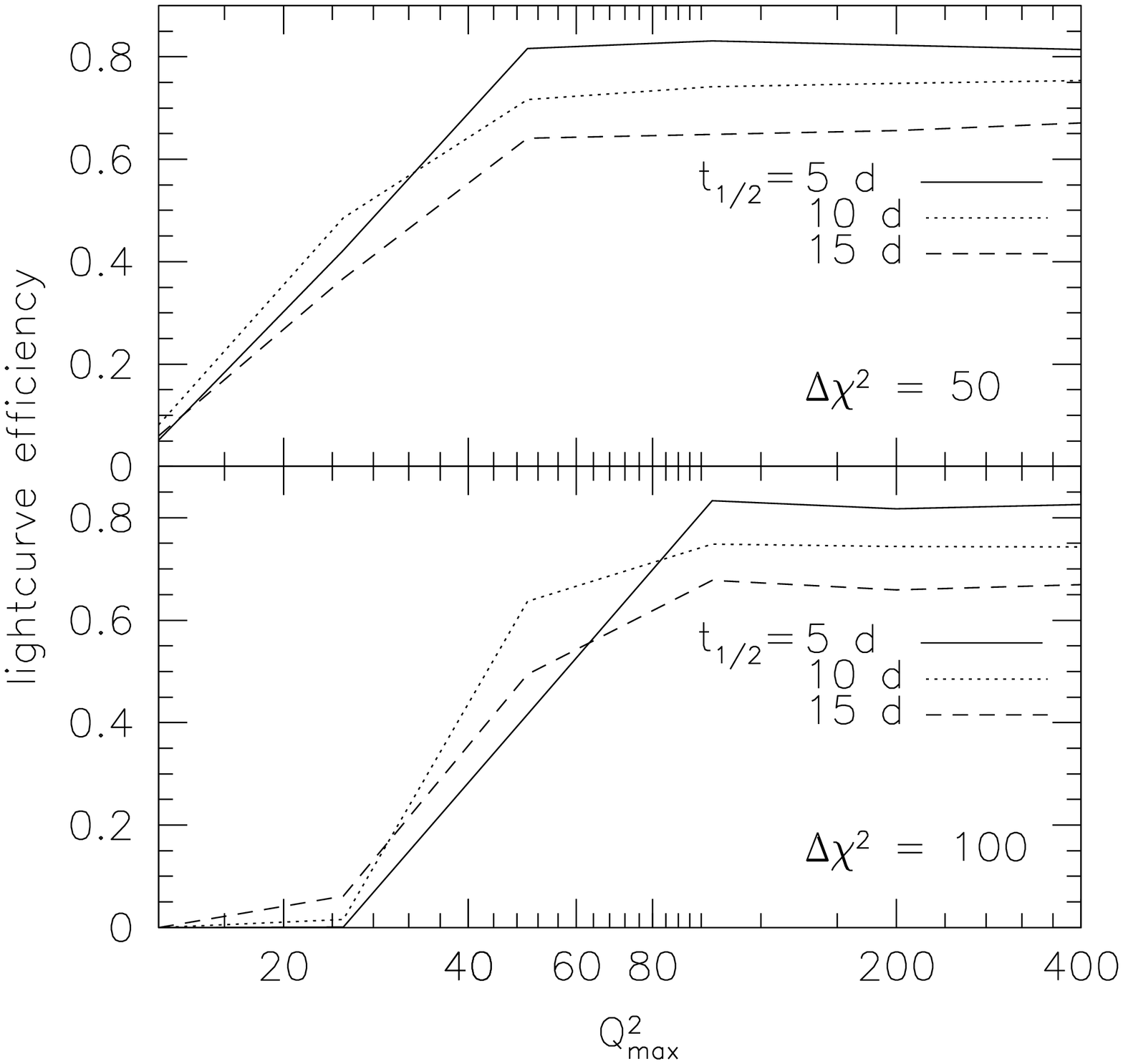}
\epsfig{width=0.49\textwidth,file=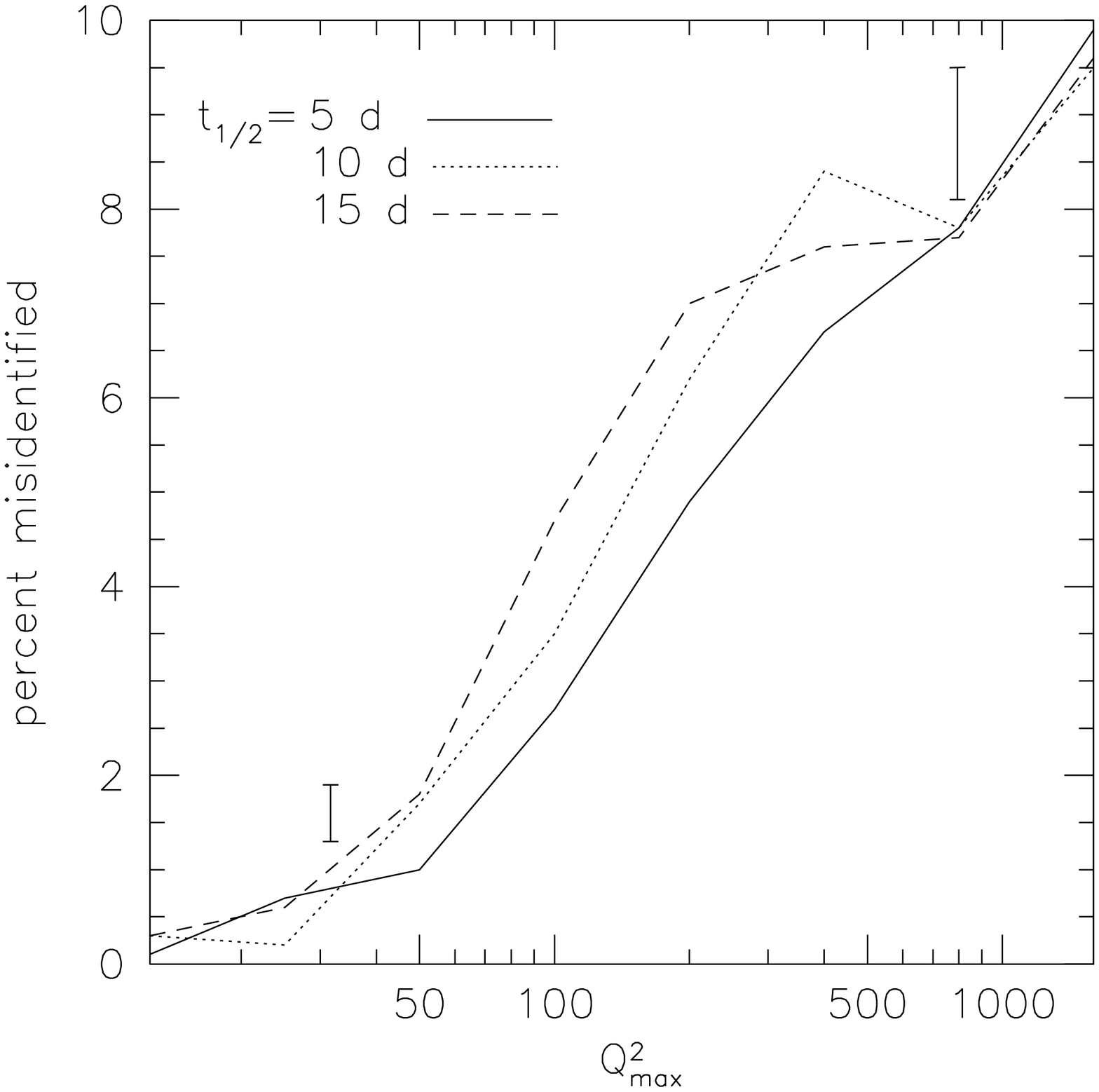}
\epsfig{width=0.49\textwidth,file=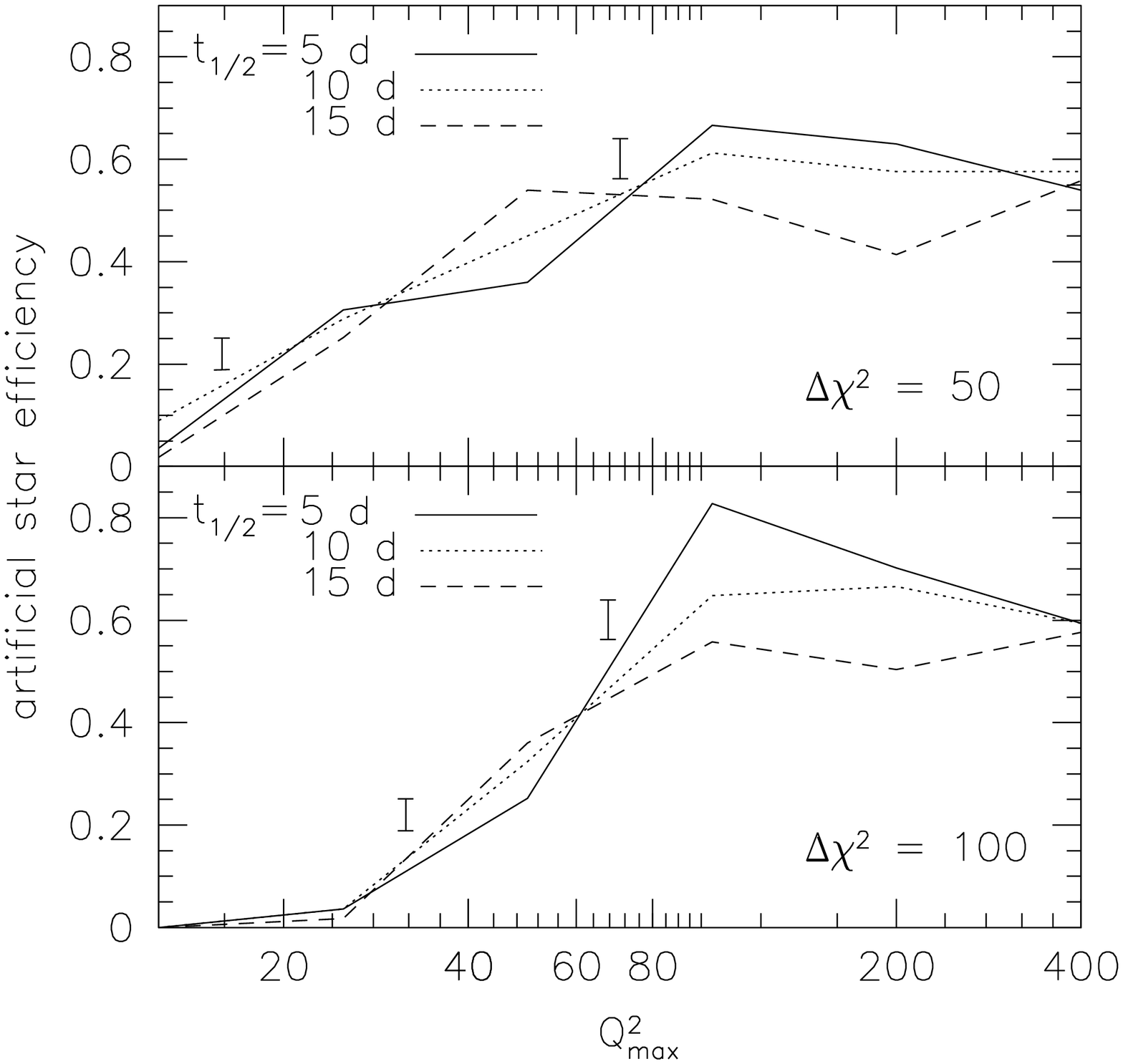}
\epsfig{width=0.49\textwidth,file=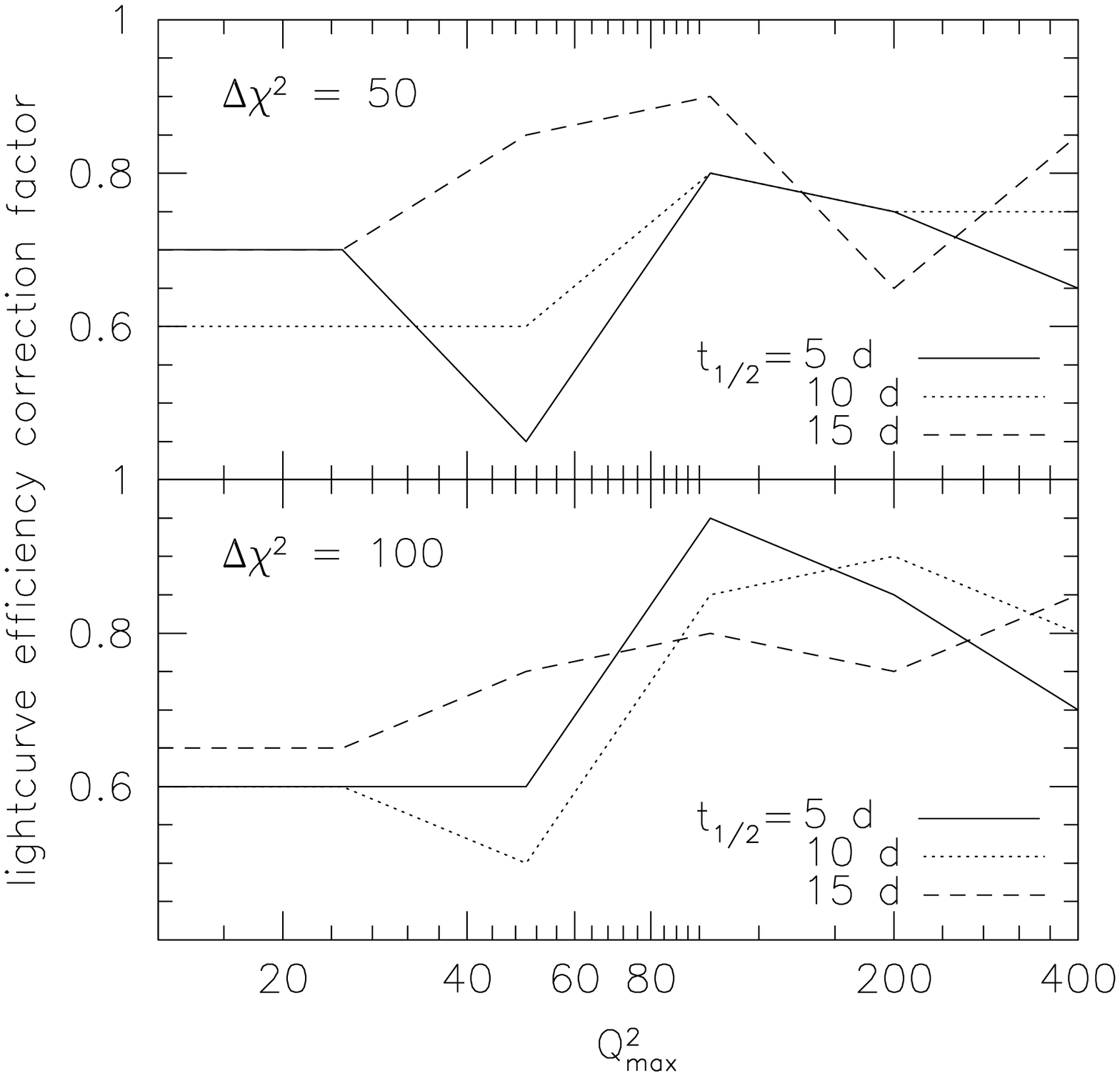}

\caption{Event detection efficiencies.  In all plots, $\qmax$ is the peak
significance (signal-to-noise) of the generated events, and $\thalf$ is their
full width at half maximum timescale.  {\em Top left:} For each entry, two
thousand artificial lightcurves were generated for comparison with the
artificial star tests.  At high significance, where a detection probability of
unity might be expected, the numbers are consistent with the $2\sigma$
requirement on $\chi^2$(fit), and the requirement that the peak be well
sampled.  Treated as a binomial distribution, the errors in the entries are
$\le 0.01$.  {\em Top right:} For each entry, two thousand artificial events
were generated for the WFC2 chip.  The hot pixel test was applied, and compared
with the results for the true WFC2 frames.  Any new hot pixels could then be
identified with the artificial events.  The misidentification probabilities are
all below 10\%.  Treated as a binomial distribution, the errors in the entries
are $\le 0.7\%$ for $10\%$ misidentification and $\le0.3\%$ for $1\%$
misidentification.  The misidentification fraction rises with peak flux as the
high gradients can fool the simple test we use.  Any high--significance events
mistakenly flagged as hot pixels would have been caught by eye.  {\em Bottom
left:} One thousand artificial events were generated, evenly divided among the
$\qmax^2$ and $\thalf$ values.  Treated as a binomial distribution, the errors
are significant: 0.039 for $p=0.5$ and decreasing to 0.031 for $p=0.2$ or 0.8
(here $p$ is the binomial probability, i.e.\ the value in the figure).  {\em
Bottom right:} This correction factor (always less than unity) is applied to
the lightcurve efficiency when calculating the microlensing rate.  It accounts
for the discrepancy between the lightcurve efficiency and the artificial star
efficiency.  These values are necessarily somewhat crude, but are adequate for
our purposes.}
\label{fig:efficiencies}
\end{figure}

From the artificial reference and difference images, the hot pixel test of
\S\ref{sec:hotpixeltest} is applied.  The probabilities of the artificial
events being identified as hot pixels are plotted in
Fig.~\ref{fig:efficiencies}.  All of these probabilities are under 10\%.

The hot pixel test is easy to apply as the optimally filtered result images are
not required.  To test the full pipeline, we proceed as follows.  One thousand
artificial events are generated similarly to the hot pixel test, with three
timescales: $\thalf=5,10,15$ days, and six peak flux levels corresponding to
$\Delta\chi^2=12.5,25,50,100,200,400$.  The artificial events are divided
roughly equally among the eighteen possibilities.  These events are now
randomly distributed in position as before.

With the artificial reference and difference images in hand, we proceed with
the full analysis of \S\ref{sec:imageanalysis} to produce optimally filtered
result images.  These artificial result images are then analyzed according to
\S\ref{sec:candidateselection}, with candidate events being selected.  This
analysis now includes the hot pixel test.  Once the lightcurves are
constructed, the analysis follows identically to \S\ref{sec:lightcurvetests}.
The artificial star efficiency so derived is plotted in
Fig.~\ref{fig:efficiencies}.

\subsection{Comparison of Efficiency Estimates}

We now compare the two efficiency calculations.  The artificial star
calculation is in principle a more accurate estimate of the efficiency, but the
lightcurve calculation is more computationally feasible.  Thus, we proceed by
checking the lightcurve efficiency with the artificial star efficiency for a
few cases, and thus deriving a correction.  The corrections to the lightcurve
efficiencies are plotted in Fig.~\ref{fig:efficiencies}.  These will
be applied to $P\left(\thalf,\qmax,\beta\right)$ and used in
\S\ref{sec:modeling} to calculate the expected rate of detectable microlensing
events.

\section{Modeling of M87 and the Virgo Cluster}
\label{sec:modeling}

To interpret the results of this search for microlensing, we must have models
for the lens populations along the line of sight to M87.  The expected rate of
microlensing events can be calculated from these models and the detection
efficiencies calculated in \S\ref{sec:detectionefficiency}.  We will need the
spatial and velocity distributions of sources (M87 stars) and lenses (M87
stars, MW halo objects, M87 halo objects, and Virgo cluster halo objects).

\subsection{Microlensing Rate}
The basic rate distribution for microlensing events can be expressed as the
integral along the the line of sight of the lens density times the cross
section (Griest 1991; Baltz \& Silk 2000):
\begin{equation}
\frac{d\Gamma_0}{d\tE}=\frac{4D_sv_c^2}{M_{\rm lens}}\int_0^1dx\, \rho_{\rm
lens}\,\omega_{\rm E}^4e^{-\omega_{\rm E}^2-\eta^2}I_0(2\omega_{\rm E}\eta),
\label{eq:simplerate}
\end{equation}
where $D_l=xD_s$, $\omega_{\rm E}=R_{\rm E}/(v_c\tE)$, $I_0$ is a modified
Bessel function of the first kind, $v_c$ is the circular velocity of the lens
population $(v_c=\sigma\sqrt{2})$, and $\eta=v_t/v_c$ is the transverse
velocity of the line of sight relative to the lens population (due to the
motion of source and observer).  This equation makes no mention of the
detection efficiency, which is added as an integral over the minimum impact
parameter $\beta$:
\begin{equation}
\frac{d\Gamma}{d\tE}=\int_0^\infty d\beta\,\frac{d\Gamma_0}{d\tE}\,
P\left[2\tE w(\beta),Q_\star\delta(\beta),\beta\right].
\end{equation}
Note that the fit parameters $\thalf$ and $\qmax$ have been replaced with more
physical ones depending on $\beta$.  Here, $Q_\star$ is the naive (photon
counting) significance with which the source star is detected if blending is
ignored.  Since $d\Gamma_0/d\tE$ is independent of $\beta$, this rate can be
written with an effective threshold value of $\beta$, determined in the obvious
manner,
\begin{equation}
\frac{d\Gamma}{d\tE}=\beta_{\rm eff}
\left(\tE,Q_\star\right)\,\frac{d\Gamma_0}{d\tE}\left(M_{\rm lens}\right),
\label{eq:rate}
\end{equation}
and the threshold $\beta_{\rm eff}$ depends only on the event timescale and
brightness of the source.  To compute the total observed rate, the distribution
is simply integrated over all $\tE$, then integrated over the mass function of
lenses and the luminosity function of sources.  Note that in our dataset $\tE$
is unknown event by event, but integrating over all $\tE$ yields the total
observed event rate.  We could have just as easily studied the rate
distribution $d\Gamma/d\thalf$ as follows:
\begin{equation}
\frac{d\Gamma}{d\thalf}=\int_0^\infty d\beta\,\frac{d\Gamma_0}{d\tE}
\left(\frac{\thalf}{2w(\beta)}\right)\,
\frac{P\left[\thalf,Q_\star\delta(\beta),\beta\right]}{2w(\beta)},
\end{equation}
but it is more expensive computationally.

This discussion of rates has neglected finite source size effects.  We include
these effects using a simple prescription (Baltz \& Silk 2000).  For a given
source flux, we can determine the required magnification to give a lensed flux
that would satisfy the requirement of the consecutivity test.  Finite source
size effects imply a maximum magnification as a function of $x$: as
$x\rightarrow 1$, the maximum magnification $\rightarrow 0$.  We can solve for
the largest $x$ allowing the required magnification, and truncate the $x$
integral in equation~\ref{eq:simplerate} accordingly.  This means thats
$d\Gamma_0/d\tE$ is now a function of $Q_\star$, both because of the required
magnification, but also in the source radius as a function of brightness.

It is interesting to point out the relationship between microlensing rate and
stellar population, first argued by Gould (1995).  All other quantities being
equal, the microlensing rate is roughly proportional to the surface brightness
fluctuation flux $\overline{F}=\langle F^2\rangle/\langle F\rangle$, where the
average is over the luminosity function of source stars.  From
equation~\ref{eq:rate}, we can argue that since the impact parameter $\beta$ is
inversely proportional to the magnification for large magnifications (the
relevant regime here), then brighter stars allow linearly proportionally larger
values of $\beta$, keeping the signal-to-noise ratio at event peak fixed.
Furthermore, these events are also observable for longer ($\thalf$ is longer),
again proportional to source flux.  Thus the rate is proportional to $F^2$,
yielding the surface brightness fluctuation flux $\overline{F}$ when integrated
over the luminosity function.  This relation is of course affected by finite
source size effects, as the maximum magnification will depend on stellar radii.

These rates assume a maxwellian velocity distribution of lenses with uniformly
moving sources.  If the source velocities are also maxwellian (as should be
approximately the case for an elliptical galaxy), the extra velocity integrals
can be separated out.  The final outcome is the same if the identification
$v_c^2({\rm lens})\rightarrow v_c^2({\rm lens})+x^2v_c^2({\rm source})$ is
made.

\subsection{Surface Brightness Profile and M87 Stars}
\label{sec:SB}
The surface brightness profile of M87 has been well studied by numerous
authors.  We will construct a composite profile from several studies that
extend from $0.02\arcsec$ to more than $150\arcsec$.  Within 20\arcsec of the
center of M87, we use the $I$-band surface brightness profile of Lauer et
al.~(1992).  We use the profile from Young et al.~(1978) to extend to
80\arcsec.  At the largest radii, out to 150\arcsec, we use the $R$-band
results of Peletier et al.~(1990).  These three measurements are spliced
together, and fit with a smoothly broken power law whose error is less than
$0.1$ mag arcsec$^{-2}$ in the range $0.1\arcsec<r<150\arcsec$, given by
\begin{equation}
\mu_I=\frac{1}{\lambda}\,\ln\left[e^{\lambda(\alpha_i\ln r+\gamma_i)}+
e^{\lambda(\alpha_o\ln r+\gamma_o)}\right]
\end{equation}
in mag arcsec$^{-2}$, where $\lambda=0.58$ governs the speed of the power law
break, $\alpha_i=0.6/\ln 10=0.26,\;\alpha_o=5.5/\ln10=2.39$ are the inner and
outer power law slopes, and $\gamma_i=15.3,\;\gamma_o=8.9$ are normalization
constants.  Note that $\lambda\rightarrow\infty$ gives a standard broken power
law, with break at $\ln r=(\gamma_o-\gamma_i)/(\alpha_i-\alpha_o)=3.01$
$(r\approx 20\arcsec)$.  At very small and very large radii, this reduces to a
single power law: $\mu_I\approx\alpha_{i,o}\ln r+\gamma_{i,o}$.

The fit to the surface brightness profile is smooth, and thus we can perform an
Abel inversion under the assumption that the system is spherically symmetric.
For convenience, we define the 2-d luminosity density
$\sigma_I=10^{-0.4\mu_I}$.  Taking its derivative, we can write down the 3-d
luminosity density,
\begin{equation}
\rho_I(r)=-2.5\log\left[-\frac{1}{\pi}\int_r^\infty\frac{d\sigma_I}{dr'}\,
\frac{dr'}{\sqrt{r'^2-r^2}}\right].
\end{equation}
This yields a luminosity density for M87, in mag arcsec$^{-3}$, which we can
apply to microlensing simulations.  The Abel inversion is done numerically,
yielding a table of density values.  This table is again fit to a smoothly
broken power law.  The fit constants are $\lambda=0.975$, $\alpha_i=2.6/\ln
10,\;\alpha_o=7.5/\ln10$, $\gamma_i=17.5,\;\gamma_o=10.9$.  This fit has errors
less than $0.075$ mag arcsec$^{-3}$ in the range $1\arcsec<r<150\arcsec$.  To
arrive at a mass density, we assume that the $I$-band mass to light ratio is
4.0 $M_\odot/L_\odot$.

We assume that the velocity distribution of M87 stars is maxwellian, with 1-d
velocity dispersion $\sigma=360$ km s$^{-1}$.  Furthermore we assume that it is
isotropic, so radial and tangential dispersions are the same.

\subsection{Milky Way Halo}

We will use simple models for the dark halos of each relevant object.  The
isothermal sphere with core has an asymptotic $1/r^2$ density profile, giving a
flat rotation curve.  The lack of observed central density cusps indicates that
a core is appropriate.  The density of lenses making up a fraction $f_{\rm MW}$
of the dark halo is given by
\begin{equation}
\rho = f_{\rm MW}\left(\frac{v_c^2}{4 \pi G}\right)\frac{1}{r^2 +r_c^2},
\end{equation}
where $v_c=220$ km s$^{-1}$ is the asymptotic rotation speed of the Milky Way.
We will take a core radius $r_c=5$ kpc, though the value doesn't much matter as
the line of sight to M87 is nearly perpendicular to the Milky Way disk.  The
velocity distribution of the halo is taken to be maxwellian, with a circular
velocity equal to $v_c$, making $\sigma=155$ km s$^{-1}$.  We impose a cutoff
 at a distance of 200 kpc from the center.

\subsection{M87 Halo}

For M87 we will also use an isothermal sphere with a core.  We will vary the
core radius, taking $r_c=5$ kpc as the fiducial value.  We assume a 1-d
velocity dispersion $\sigma=360$ km s$^{-1}$, giving $v_c=510$ km s$^{-1}$.
Based simply on dispersion velocity, M87 is roughly five times as massive as
the Milky Way.  We allow the halo lens fraction $f_{\rm M87}$ to be
independent of $f_{\rm MW}$.

\subsection{Virgo Cluster Halo}

The halo of the Virgo cluster is more problematic.  Again, we will assume an
isothermal sphere with a core, with $\sigma=1000$ km s$^{-1}$ ($v_c=1400$ km
s$^{-1}$).  We will assume that M87 is centered in the Virgo halo.  This
approach requires that we assign a very large core radius to the Virgo halo:
for $r_c$ as small as 100 kpc, Virgo dominates at a radius of 40 kpc from the
center of M87.  Again, we allow an independent halo lens fraction $f_{\rm
Vir}$.

\subsection{Expected Event Rates}

Armed with the model for M87 and the Virgo cluster, we can now proceed to
compute the expected rate of detectable microlensing events.  Some experimental
parameters are required.  Furthermore, we must make assumptions about the
luminosity function of M87 stars, and about the mass function of the lenses.

The capabilities of the WFPC2 can be summarized for our purposes as follows.
We assume that the zero point in F814W is $m_I^{\rm ZP}=23.86$ mag in the
$I$-band (this flux gives one photo-electron per second), and that only the
F814W frames are used to detect events: namely an exposure of $t_{\rm
obs}=4\times 260$ s = 1040 s per orbit.  The zero point for this exposure time
is $m_I^0=m_I^{\rm ZP}+2.5\,\log 1040=31.40$ mag, giving one photo-electron
over the exposure.  The resolution can be characterized by one number per chip,
$\Omega_{\rm PSF} = 1/\sum \psi_i^2$, where $\psi_i$ is the normalized PSF
(Gould 1996).  Measured in pixels (taking (0.0455~arcsec)$^2$ for PC1 and
(0.1~arcsec)$^2$ for WFC2, WFC3, WFC4), the PSF sizes are 20.0, 7.31, 9.05,
7.89 for the PC1, WFC2, WFC3, and WFC4, respectively.  These are somewhat less
than 0.1 arcsec$^2$.  Lastly, we assume that the read noise is 5
photoelectrons, and that the dark noise is negligible.

We assume that the source population is circularly symmetric, with surface
brightness taken from \S\ref{sec:SB}.  We take the background sky brightness to
be a uniform $\mu_{\rm sky}=21.5$ mag arcsec$^{-2}$ in the $I$-band.  We will
compute the microlensing rate at radial positions spaced by 1 arcsec, starting
1 arcsec from the center of M87 and extending to 200 arcsec.  The
two-dimensional microlensing rate is trivially constructed from this.

The surface brightness serves to normalize the luminosity function for stars in
M87.  We assume that M87 has the same luminosity function found for the
Galactic bulge by Terndrup, Frogel \& Whitford (1990).  We adjust the
high-luminosity cutoff to give a surface brightness fluctuation magnitude of
$\overline{M}_I=-1.5$, appropriate for M87.  Taking a distance modulus of
$D=31$ to the Virgo cluster, we can express the signal-to-noise ratio to detect
a star of absolute magnitude $M_I$ magnified by a factor $A=1+\delta$,
\begin{equation}
Q=\frac{10^{-0.4\left(M_I+D+m_I^0/2\right)}\;\delta}
{\sqrt{\Omega_{\rm PSF}\left(10^{-0.4\mu_I}+10^{-0.4\mu_{\rm sky}}\right)}}.
\end{equation}

Lastly, we fix the mass functions for both the stellar component of M87, and
for the lenses.  We take the Chabrier (2001) mass function for stars, which has
an effective peak around 0.1 $M_\odot$.  For the lenses, we take a delta
function at $-1/2$ dex solar (0.32$M_\odot$), similar to the best value found
by the MACHO collaboration (Alcock et al.~2000).

We now have a complete model for calculating the microlensing event rate.  We
use thresholds of $\Delta\chi^2=50$ and 100, and we vary the core radii of both
the M87 and Virgo cluster halos.  The results are summarized in
Table~\ref{tab:rates}, clearly showing the dependence on core radii and on lens
fractions $f_{\rm MW}$, $f_{\rm M87}$ and $f_{\rm Vir}$.

From these simulations of microlensing in this dataset, we conclude that of
order 10 events are expected for $f=1$.  Taking $f=0.2$ as indicated by MACHO
(Alcock et al.~2000), we expect 1--2 events with the sensitivity to
microlensing that we were able to achieve, namely $\Delta\chi^2>100$.

\section{Discussion}
\label{sec:discussion}

We have identified seven candidate variable point sources in M87.  The obvious
question is what are these sources.  We will discuss several possibilities
below.  If any of these candidates are in fact due to microlensing, we want to
understand the implications for populations of lenses associated with the Virgo
Cluster.

Perhaps most obviously, any of these candidate events could potentially be
classical novae, with the exception of PC1-3 which is far too red.  A study of
these candidates as novae will be reported elsewhere (Shara et al.~2003).  How
many novae might we expect to see in M87 during a 30 day run?  A simple, purely
theoretical estimate is as follows. The space density of cataclysmic variables
(CVs) near the Sun is roughly $10^{-4}$ that of all stars.  All CVs undergo
thermonuclear runaways -- nova eruptions -- when their white dwarf accretes
enough hydrogen-rich matter (typically $10^{-5} M_\odot$) from the main
sequence companion. The accretion timescale (and hence inter-eruption
timescale) is often of order $10^5$ to $10^6$ years. If the CVs in M87 are
similar to those in the solar neighborhood, then there should be $\sim 10^8$
CVs among the $\sim 10^{12}$ stars of M87.  Thus we expect 100-1000 nova
eruptions/year in M87, or $\sim$ 10-100 nova eruptions in M87 during a
month-long survey. As we are almost certainly {\em not} complete in our
detections of low luminosity novae, or those located close to the galaxy's
nucleus, the $\sim 6$ likely/possible novae we do observe are in good agreement
with the simple prediction.

Candidate PC1-3 is likely to be a Mira variable.  From the lightcurve it
appears to vary by at least 2 magnitudes, but Miras can exhibit variations much
larger than this.  Candidate PC1-4 is our second reddest, but it is fairly blue
to be a Mira.  However, Miras are known to be bluer at maximum light (Kanbur,
Hendry \& Clarke 1997), so it is not unreasonable to suppose this might also be
a Mira.

From its shape, candidate WFC2-5 is an excellent microlensing candidate.  In
addition, its color is quite constant throughout the time series.  However, we
can not rule out the possibility that it is a nova.  Its blue color is
certainly consistent with the nova hypothesis.  In fact for it to be
microlensing the source would have to be e.g.\ a horizontal branch star.  On
numbers alone, we expect that most microlensing events will be red giants, so
this is puzzling.  Since the horizontal branch lies at $M_I\approx0.25$ for
$V-I=0.35$, the implied magnification for WFC2-5 is roughly 620.  From the
full-width at half maximum timescale $\thalf=7$ days, the implied Einstein time
is $\tE\approx 2500$ days.  The peak of the distribution $d\Gamma/d\log\tE$ is
at roughly 75 days for typical stellar mass lenses.  This implies that a source
3.8 magnitudes brighter than the horizontal branch is typical.  This is near
the tip of the red giant branch.  In other words, we expect most events to be
much lower magnifications of much brighter stars.  This seems to indicate that
the horizontal branch microlensing hypothesis is disfavored.  We note that this
source is much brighter than the aperiodic blue variables, otherwise known as
blue bumpers, observed by the MACHO collaboration (Keller et al.~2002).  Such
sources vary by less than 0.5 $V$ magnitudes, at $M_V\sim-3$.

We have performed a simple test for the presence of finite source effects in
candidate WFC2-5.  Following Yoo et al.~(2004), we introduce one more fit
parameter, the angular size of the source relative to the Einstein angle:
$\rho=\theta_\star/\theta_{\rm E}$.  With impact parameter $u$ in Einstein
units as before, we define $z=u(t)/\rho$, $\zeta=\beta/\rho$, and thus
$z=(u(t)/\beta)\zeta$.  The degenerate microlensing lightcurve with finite
source effects is now
\begin{equation}
F(t)=B+\frac{\Dfm\beta}{u(t)}\left(\frac{4}{\pi}\right)z\,
E\left(\min\left(z^{-1},1\right),z\right),
\end{equation}
with $\zeta$ being the new fit parameter (since $u(t)/\beta$ is already fit for
with $t_0$ and $\thalf$; see equations \ref{eq:lc1} and \ref{eq:lc2}), and for
simplicity we assume no limb darkening.  Note that $E$ is the elliptic integral
of the second kind.  We find a new best fit, with $\Dfm=-10.01$ magnitudes,
$\thalf=0.28$ days, $t_0=23.31$ days, and $\zeta=0.0336$.  This fit implies a
much larger (20.5x) naive magnification, and much shorter (25.1x) naive
timescale.  The minimum impact parameter is roughly 1/30 of the stellar radius,
namely the finite source effects are severe.  The fit is not overwhelmingly
better; it is slightly wider and flatter near the peak. A simple $F$-test
indicates that finite source effects exist at 87\% confidence.  However, the
higher magnification implied by the finite source fit is less likely by a
factor of 20.  For a horizontal branch star, $A=1/\beta=1.2\times 10^4$,
implying $\rho=2.5\times10^{-3}$.  Taking $R_\star=5R_\odot$, and a solar mass
lens, $D_{\rm ls}=10$ kpc.  This is reasonable for an M87 halo lens, but
probably not for an M87 star.  Returning to the fit without finite source size
effects, $\beta=1.6\times10^{-3}$ for the horizontal branch source.  Starting
with this fit, and computing $\chi^2$ as a function of $\zeta$, we find that
$\chi^2$ is pretty flat as a function of $\zeta$ for $\zeta>1$, but that it
blows up for $\zeta<0.7$ (by this we mean that $\Delta\chi^2$ goes from 2 at
$\zeta=0.69$ to 5 at $\zeta=0.65$).  Enforcing the condition that $\zeta>0.7$
for the horizontal branch star requires that $\rho<2.3\times10^{-3}$, implying
$D_{\rm ls}>13$ kpc for a solar mass lens.  Again, this would require an M87
halo or Virgo halo lens.

Considering the microlensing hypothesis, we would expect to detect 1-2
microlensing events from a 20\% microlensing halo for the Virgo cluster with
the sensitivity level achieved.  Having one solid candidate is certainly
consistent with that, though even in the case of zero candidates, the limits we
might place on the Virgo lens fraction are not strong: naively the 95\%
confidence limit on the lens fraction is $f_{\rm Vir}<0.6$.

We have shown that it is possible to detect variables near the photon noise
limit with repeat observations using HST.  In the future, a continuation of
this work using the much more sensitive Advanced Camera for Surveys (ACS) would
allow a huge increase in sensitivity to microlensing.  Firstly, the area
covered is twice as large, second the efficiency is 4.5 times higher in the
$I$-band, and third, $\Omega_{\rm PSF}$ is a factor of 1.6 smaller.  These
factors combined allow a factor of {\em fourteen} increase in sensitivity for
the same time coverage.  Clearly this is a huge advantage, meaning a 20\% Virgo
halo would contribute more like 15-30 events in a one-month program.  In
addition, we believe that the sensitivity could be made significantly higher by
altering the pointing by several PSF diameters from visit to visit, allowing a
complete decoherence between source and detector structure, thus removing
essentially all hot pixels from the type of variability search we performed.
Allowing a lower threshold would obviously be a significant improvement.

We have reported on microlensing candidates observed toward M87.  We have shown
that the HST is a powerful tool for this kind of science.  The improvements
that would be allowed by the ACS are striking, and would definitively detect,
or rule out at high confidence, a microlensing halo around the Virgo cluster.

\acknowledgments

E.B. wishes to thank D.~Alves and A.~Crotts for useful conversations, and
acknowledges support from the Columbia University Academic Quality Fund.  We
thank the referee for thorough and insightful comments that improved this paper
significantly.  We gratefully acknowledge support under grant \mbox{GO-8592}
from the Space Telescope Science Institute.  This research is based on
observations made with the NASA/ESA Hubble Space Telescope obtained at the
Space Telescope Science Institute.  STScI is operated by the Association of
Universities for Research in Astronomy, Inc.\ under NASA contract NAS 5-26555.

\begin{deluxetable}{rrrrrrr}
\tablewidth{0pt}
\tablecaption{Journal of Observations\label{tab:journal}}
\tablehead{\multicolumn{1}{c}{Visit} & \multicolumn{1}{c}{Date} & 
\multicolumn{1}{c}{Date} & \multicolumn{1}{c}{$\Delta T$}&
\multicolumn{1}{r}{$\Delta x$} & \multicolumn{1}{r}{$\Delta y$} &
\multicolumn{1}{c}{Dataset}\\
& \multicolumn{1}{c}{(GMT)} & \multicolumn{1}{c}{(MJD)} &
\multicolumn{1}{c}{(days)} &
\multicolumn{2}{c}{($0.05\arcsec$ pixels)}}
\startdata
 1&  May 28 2001&52057.430729& 0.00& 0.00& 0.00&U6730101R-5R\\
 2&  May 29 2001&52058.366840& 0.94& 0.05& 0.08&U6730201R-5R\\
 3&  May 30 2001&52059.303645& 1.87& 1.07& 0.10&U6730301R-5R\\
 4&  May 31 2001&52060.239756& 2.81& 0.93& 0.56&U6730401R-5R\\
 5& June  1 2001&52061.243229& 3.81& 1.02&-0.67&U6730501R-5R\\
 6& June  2 2001&52062.313367& 4.88& 1.01&-1.02&U6730601R-5R\\
 7& June  3 2001&52063.383506& 5.95& 0.01&-0.88&U6730701R-5R\\
 8& June  4 2001&52064.453645& 7.02&-0.31& 0.40&U6730801R-5R\\
 9& June  5 2001&52065.389756& 7.96&-0.05& 0.36&U6730901R-5R\\
10& June  6 2001&52066.326562& 8.90& 0.05& 0.03&U6731001R-5R\\
11& June  7 2001&52067.262673& 9.83& 1.06& 0.03&U6731101R-5R\\
12& June  8 2001&52068.333506&10.90& 1.15& 0.59&U6731201R-5R\\
13& June  9 2001&52069.269618&11.84& 1.05&-0.50&U6731301R-5R\\
14& June 10 2001&52070.210590&12.78& 1.12&-0.61&U6731401R-5R\\
15& June 11 2001&52071.145312&13.71& 0.27&-1.40&U6731501R-5R\\
16& June 12 2001&52072.147395&14.72& 0.16&-1.16&U6731601R-5R\\
17& June 13 2001&52073.218229&15.79&-0.02& 0.34&U6731701R-5R\\
18& June 14 2001&52074.152951&16.72& 0.01& 0.21&U6731801R-5R\\
19& June 15 2001&52075.093229&17.61& 1.07& 0.24&U6731901R-5R\\
20& June 16 2001&52076.027951&18.60& 1.09& 0.11&U6732001R-5R\\
21& June 16 2001&52076.961979&19.43& 1.10&-0.76&U6732101R-5R\\
22& June 18 2001&52078.032812&20.60& 1.13&-0.80&U6732201R-5R\\
23& June 19 2001&52079.103645&21.67&-0.01&-0.76&U6732301R-5R\\
24& June 20 2001&52080.039062&22.61& 0.17&-0.96&U6732401R-5R\\
25& June 21 2001&52080.978646&23.61& 0.03& 0.44&U6732501R-5R\\
26& June 22 2001&52081.982118&24.61& 0.01& 0.38&U6732601R-5R\\
27& June 23 2001&52082.984896&25.62& 1.24&-0.06&U6732701R-5R\\
28& June 24 2001&52084.052257&26.62& 1.16& 0.05&U6732801R-5R\\
29& June 25 2001&52085.055034&27.62& 1.06&-0.69&U6732901R-5R\\
30& June 25 2001&52085.993923&28.56& 1.32&-1.16&U6733001R-5R\\
\enddata

\tablecomments{The date refers to that of the first observation
in a given visit. Each visit comprises four dithered F814W images,
followed by a single F606W image.  Offsets are shown for CCD WFC2 only
in units of $0\farcs05$ subpixels relative to the origin of the
first visit.}
\end{deluxetable}

\onecolumn

\begin{deluxetable}{lrrcrrcccl}
\tablewidth{0pt}
\tablecaption{Candidate Events\label{tab:candidate}}

\tablecomments{Final candidate list passing all cuts.  Pixel coordinates are
given, both as the center pixel of the group that passes all cuts, and as a
flux--weighted centroid.  One event is an excellent microlensing candidate,
well sampled on both sides of the peak.  Two candidates are obvious novae.  The
remainder are probably variable stars.  Note that only the PC1 and the WFC2
chips had candidates passing all cuts.}

\tablehead{
number & \multicolumn{2}{c}{pixel (fit)} & & \multicolumn{2}{c}{pixel (flux)} &
radius & $\alpha$ & $\delta$ &  comments\\
 & \multicolumn{1}{c}{$x$} & \multicolumn{1}{c}{$y$} & &
\multicolumn{1}{c}{$x$} & \multicolumn{1}{c}{$y$} & (arcsec) &
(J2000) & (J2000)}

\startdata
PC1-1 & 530 & 439 & & 528.6 & 438.6 & 3.5 & 12 30 49.744 & 12 23 29.39 &
classical nova\\
PC1-2 & 102 & 216 & & 101.0 & 214.6 & 17.4 & 12 30 48.285 & 12 23 31.13 &
rising\\
PC1-3 & 586 & 742 & & 585.3 & 741.5 & 16.4 & 12 30 50.262 & 12 23 17.73 &
declining\\
PC1-4 & 767 & 267 & & 766.8 & 265.9 & 17.4 & 12 30 50.204 & 12 23 40.88 &
declining\\
\tableline
WFC2-5 & 507 & 255 & & 505.9 & 254.4 & 49.2 & 12 30 47.684 & 12 22 46.32 &
microlensing candidate\\
WFC2-6 &  94 & 449 & &  93.6 & 448.4 & 61.4 & 12 30 45.408 & 12 23 16.17 &
globular cluster nova\\
WFC2-7 & 788 & 399 & & 787.6 & 398.5 & 78.3 & 12 30 47.557 & 12 22 14.89 &
declining\\
\enddata
\end{deluxetable}

\begin{deluxetable}{lccrrccl}
\tablewidth{0pt}
\tablecaption{Candidate Event Fit Parameters\label{tab:candidatefits}}

\tablecomments{Microlensing fit parameters for final candidates.  These are the
maximum flux increase expressed in absolute magnitude (taking $D=31$), the
full-width at half maximum $\thalf$, peak time $t_0$ as $\Delta$MJD = Modified
Julian Date $-$ 52057, frame with maximum flux (1-30), $V-I$ color of the {\em
excess} flux (obtained with aperture photometry), and the goodness of the
microlensing fit.  For WFC2-5, the only event with both good coverage of the
peak and a good microlensing fit, the fit errors are: $\Dfm=-6.73\pm0.08$
magnitudes, $\thalf=7.02\pm1.40$ days, $t_0=23.26\pm0.26$ days.  Errors on the
fit parameters of the other events are much less meaningful.}

\tablehead{number & $\Dfm$ & $\thalf$ &
 \multicolumn{1}{c}{$t_0$} & \multicolumn{1}{c}{peak} & $V-I$ & $\chi^2/$dof &
 comments\\
 & $(M_I)$ & (days) & ($\Delta$MJD) & (frame) & & & }

\startdata
PC1-1 & -8.85 & 1.36 & 23.58 & 24 &
$0.29\pm0.12$ &  3.18 & classical nova\\
PC1-2 & -7.41 & 16.4 & 26.15 & 28 &
$0.31\pm0.07$ & 2.13 & rising\\
PC1-3 & -7.97 & 19.7 & 2.94 & 3 &
$1.21\pm0.07$ & 1.02 & declining\\
PC1-4 & -7.37 & 7.11 & -0.28 & 1 &
$0.66\pm0.14$ & 0.74 & declining\\
\tableline
WFC2-5 & -6.73 & 7.02 & 23.26 & 24 &
$0.35\pm0.12$ & 1.23 & microlensing candidate\\
WFC2-6 & -8.09 & 0.91 & 11.58 & 12 &
$0.25\pm0.17$ & 3.67 & globular cluster nova\\
WFC2-7 & -6.49 & 11.6 & -0.69 & 2 &
$0.31\pm0.21$ & 1.43 & declining\\
\enddata
\end{deluxetable}

\begin{deluxetable}{lllllll}
\tablewidth{0pt} \tablecaption{Expected Microlensing Rate
\label{tab:rates}}

\tablecomments{Expected number of microlensing events for each component of the
model.  The self lensing component is quite small, less than 0.25 events
expected.  The dominant component is clearly the Virgo cluster halo.  For
$f_{\rm M87}=0.2$, the M87 halo contribution is comparable to the self lensing.
The Milky Way halo contribution is quite small, and with $f_{\rm MW}=0.2$, it
is much less than even the self lensing component.}

\tablehead{
& \multicolumn{1}{c}{threshold} & &
\multicolumn{1}{l}{$\Delta\chi^2=50$} & &
\multicolumn{1}{l}{$\Delta\chi^2=100$}}
\startdata
M87 stars & & & 0.55 & & 0.22\\ \tableline
Milky Way Halo & & & 0.32 $f_{\rm MW}$ & & 0.19 $f_{\rm MW}$\\ \tableline
M87 Halo & $r_c=2$ kpc & & 2.65 $f_{\rm M87}$ & & 1.32 $f_{\rm M87}$\\
 & $r_c=5$ kpc & & 2.38 $f_{\rm M87}$ & & 1.21 $f_{\rm M87}$\\
 & $r_c=10$ kpc & & 2.02 $f_{\rm M87}$ & & 1.06 $f_{\rm M87}$\\ \tableline
Virgo Halo & $r_c=100$ kpc & & 14.2 $f_{\rm Vir}$ & & 8.03 $f_{\rm Vir}$ \\
 & $r_c=200$ kpc & & 10.5 $f_{\rm Vir}$  & & 6.13 $f_{\rm Vir}$ \\
 & $r_c=500$ kpc & & 5.91 $f_{\rm Vir}$ & & 3.60 $f_{\rm Vir}$ \\ \tableline
Totals & & &(9---18) $f$ & &(5---10) $f$
\enddata
\end{deluxetable}

\end{document}